\documentclass[12pt]{article}
\usepackage[dvips]{graphicx}

\begin {document}
\begin{flushright}
{\small
SLAC--PUB--10871\\
November 2004\\}
\end{flushright}

\def\pl{{\it Phys Lett\/}}
\newcommand{\gsim} {\buildrel > \over {_\sim}}
\newcommand{\eg}   {{\em e.g.}}
\newcommand{\ket}[1]{\,\left|\,{#1}\right\rangle}
\newcommand{\cd}{\makebox[0.08cm]{$\cdot$}}
\newcommand{\MSbar} {\hbox{$\overline{\hbox{\tiny MS}}$}}
\newcommand{\M}{\mathcal{M}}
\newcommand{\qu}{{\rm q}}
\newcommand{\qb}{${\rm\bar q}$}
\newcommand{\pvec}{\vec p}
\newcommand{\kvec}{\vec k}
\newcommand{\rvec}{\vec r}
\newcommand{\Rvec}{\vec R}
\newcommand{\ieps}{i\varepsilon}
\newcommand{\order}[1]{${ O}\left(#1 \right)$}
\renewcommand{\bar}[1]{\overline{#1}}
\newcommand{\half}{{\frac{1}{2}}}
\newcommand{\threehalf}{{\frac{3}{2}}}

\vfill

\begin{center}
{{\bf\LARGE Light-Front QCD}\footnote{Work supported by Department
of Energy contract DE--AC02--76SF00515.}}

\bigskip
{\it Stanley J. Brodsky \\
Stanford Linear Accelerator Center \\
Stanford University, Stanford, California 94309 \\
E-mail:  sjbth@slac.stanford.edu}
\medskip
\vfill

 Invited lectures and talk presented at the\\
58th Scottish University Summer School In Physics: A NATO Advanced
Study Institute and the European Union Hadronic Physics 13 Summer
Institute (SUSSP58)\\
  St. Andrews, Scotland\\
 30 August--1 September 2004

 \vfill
 \end{center}

\newpage

$$ $$

\begin{center}
{\bf\large Abstract }
\end{center}
In these lectures, I survey a number of applications of light-front
methods to hadron and nuclear physics phenomenology and dynamics,
including light-front statistical physics. Light-front Fock-state
wavefunctions provide a frame-independent representation of hadrons
in terms of their fundamental quark and gluon degrees of freedom.
Nonperturbative methods for computing LFWFs in QCD are discussed,
including string/gauge duality which predicts the power-law fall-off
at high momentum transfer of light-front Fock-state hadronic
wavefunctions with an arbitrary number of constituents and orbital
angular momentum. The AdS/CFT correspondence has important
implications for hadron phenomenology in the conformal limit,
including an all-orders derivation of counting rules for exclusive
processes. One can also compute the hadronic spectrum of
near-conformal QCD assuming a truncated AdS/CFT space. Given the
LFWFs, one can compute form factors, heavy hadron decay amplitudes,
hadron distribution amplitudes, and the generalized parton
distributions underlying deeply virtual Compton scattering. The
quantum fluctuations represented by the light-front Fock expansion
leads to novel QCD phenomena such as color transparency, intrinsic
heavy quark distributions, diffractive dissociation, and
hidden-color components of nuclear wavefunctions. A new test of
hidden color in deuteron photodisintegration is proposed. The origin
of leading-twist phenomena such as the diffractive component of deep
inelastic scattering, single-spin asymmetries, nuclear shadowing and
antishadowing is also discussed; these phenomena cannot be described
by light-front wavefunctions of the target computed in isolation.
Part of the anomalous NuTeV results for the weak mixing angle
$\theta_W$ could be due to the non-universality of nuclear
antishadowing for charged and neutral currents.

\newpage

\section{Introduction}

In principle, quantum chromodynamics provides a fundamental
description of hadronic and nuclear structure and dynamics in terms
of their elementary quark and gluon degrees of freedom. The theory
has extraordinary properties such as color
confinement~\cite{Greensite:2003bk}, asymptotic
freedom~\cite{Gross:1973id,Politzer:1973fx}, a complex vacuum
structure, and it predicts an array of new forms of hadronic matter
such as gluonium and hybrid states~\cite{Klempt:2000ud}. The phase
structure of QCD~\cite{Rajagopal:2000wf} implies the formation of a
quark-gluon plasma in high energy heavy-ion
collisions~\cite{Rischke:2003mt} as well insight into the evolution
of the early universe~\cite{Schwarz:2003du}. Its non-Abelian Yang
Mills gauge theory structure provides the foundation for the
electroweak interactions and the eventual unification of the
electrodynamic, weak, and hadronic forces at very short distances.

The asymptotic freedom property of QCD explains why the strong
interactions become weak at short distances, thus allowing hard
processes to be interpreted directly in terms of the perturbative
interactions of quark and gluon quanta. This in turn leads  to
factorization theorems~\cite{Collins:1989gx,Bodwin:1984hc} for
both inclusive and exclusive processes~\cite{Brodsky:1989pv} which
separate the hard scattering subprocesses which control the
reaction from the nonperturbative physics of the interacting
hadrons.

QCD becomes scale free and conformally symmetric in the analytic
limit of zero quark mass and zero $\beta$ function.   Conversely,
one can start with the conformal prediction and systematically
incorporate the non-zero $ \beta$ function contributions into the
scale of the running coupling. This ``conformal correspondence
principle" determines the form of the expansion polynomials for
distribution amplitudes and the behavior of nonperturbative
wavefunctions which control hard exclusive processes at leading
twist.  The conformal template also can be used to derive
commensurate scale relations which connect observables in QCD
without scale or scheme ambiguity.

Recently, a remarkable duality has been established between
supergravity string theory in 10 dimensions and conformal
supersymmetric extensions of
QCD~\cite{Maldacena:1997re,Polchinski:2001tt,Brower:2002er,Andreev:2002aw}.
The AdS/CFT correspondence is now leading to a new understanding of
QCD at strong coupling and the implications of its near-conformal
structure.  As I will discuss here, the AdS/CFT correspondence of
large $N_C$ supergravity theory in higher-dimensional  anti-de
Sitter space with supersymmetric QCD in 4-dimensional space-time has
important implications for hadron phenomenology in the conformal
limit, including the nonperturbative derivation of counting rules
for exclusive processes and the behavior of structure functions at
large $x_{bj}.$  String/gauge duality also predicts the QCD
power-law fall-off of light-front Fock-state hadronic wavefunctions
with arbitrary orbital angular momentum at high momentum transfer.

The  Lagrangian density of QCD~\cite{Fritzsch:1973pi} has a deceptively
simple form:
\begin{equation}
{\cal L}=   {\overline \psi}(i\gamma_\mu D^\mu-m)\psi
-{1\over 4} G^2_{\mu \nu}
\end{equation}
where the covariant derivative is $i D_\mu = i \partial_\mu - g
A_\mu$ and where the gluon field strength is $G_{\mu \nu} = {i\over
g}[D_\mu, D_\nu]$. The structure of the QCD Lagrangian is dictated
by two principles: (i) local $SU(N_C)$ color gauge invariance -- the
theory is invariant when a quark field is rotated in color space and
transformed in phase by an arbitrary unitary matrix $\psi(x) \to
U(x) \psi(x)$ locally at any point $x^\mu$ in space and time;  and
(ii) renormalizability, which requires the appearance of dimension-four
interactions. In principle, the only parameters of QCD are the
quark masses and the QCD coupling determined from a single
observable at a single scale.

Solving QCD from first principles is extremely challenging because
of the non-Abelian three-point and four-point gluonic couplings
contained in its Lagrangian. The analytic problem of describing QCD
bound states is compounded not only by the physics of confinement,
but also by the fact that the wavefunction of a composite of
relativistic constituents has to describe systems of an arbitrary
number of quanta with arbitrary momenta and helicities.  The
conventional Fock state expansion based on equal-time quantization
quickly becomes intractable because of the complexity of the vacuum
in a relativistic quantum field theory. Furthermore, boosting such a
wavefunction from the hadron's rest frame to a moving frame is as
complex a problem as solving the bound state problem itself.  The
Bethe-Salpeter bound state formalism, although manifestly covariant,
requires an infinite number of irreducible kernels to compute the
matrix element of the electromagnetic current even in the limit
where one constituent is heavy.

The description of relativistic composite systems using light-front
quantization is in contrast remarkably simple.  The Heisenberg
problem for QCD can be written in the form
\begin{equation}
H_{LC }\vert H\rangle = M_H^2 \vert H\rangle\; ,
\end{equation}
where $H_{LC}=P^+ P^- - P_\perp^2$ is the mass operator.  The
operator $P^-=P^0-P^3$ is the generator of translations in the
light-front time $x^+=x^0+x^3.$   Its form is predicted
from the QCD Lagrangian. The quantities $P^+=P^0+P^3$ and
$P_\perp$ play the role of the conserved three-momentum.
The simplicity of the light-front Fock representation relative to
that in equal-time quantization arises from the fact that the
physical vacuum state has a much simpler structure on the light
cone.  Indeed, kinematical arguments suggest that the light-front
Fock vacuum is the physical vacuum state.  This means that all
constituents in a physical eigenstate are directly related to that
state, and not disconnected vacuum fluctuations.  In the light-front
formalism the parton model is literally true.

Formally, the light-front expansion is constructed by quantizing QCD
at fixed light-front time \cite{Dirac:1949cp} $\tau = t + z/c$ and
forming the invariant light-front Hamiltonian: $ H^{QCD}_{LF} = P^+
P^- - {\vec P_\perp}^2$ where $P^\pm = P^0 \pm
P^z$~\cite{Brodsky:1997de}.  The momentum generators $P^+$ and $\vec
P_\perp$ are kinematical; {\em i.e.}, they are independent of the
interactions. The generator $P^- = i {d\over d\tau}$ generates
light-front time translations, and the eigen-spectrum of the Lorentz
scalar $ H^{QCD}_{LF}$ gives the mass spectrum of the color-singlet
hadron states in QCD together with their respective light-front
wavefunctions.

Each hadronic eigenstate $\vert H\rangle$ of the QCD light-front
Hamiltonian can be expanded on the complete set of eigenstates
$\{\vert n\rangle\} $ of the free Hamiltonian which have the same
global quantum numbers: $\vert H\rangle=\sum\psi^H_n(x_i, k_{\perp},
\lambda_i) \vert n\rangle.$ For example, the proton state satisfies:
$ H^{QCD}_{LF} \ket{\psi_p} = M^2_p \ket{\psi_p}$. This equation can
be written as a Heisenberg matrix eigenvalue problem by introducing
a complete set of free Fock states.  The Fock expansion begins with
the color singlet state $\vert u u d \rangle $ of free quarks, and
continues with $\vert u u d g \rangle $ and the other quark and
gluon states that span the degrees of freedom of the proton in QCD.
The Fock states $\{\vert n\rangle \}$ are built on the free vacuum
by applying the free light-front creation operators. The summation
is over all momenta $(x_i, k_{\perp i})$ and helicities $\lambda_i$
satisfying momentum conservation $\sum^n_i x_i = 1$ and $\sum^n_i
k_{\perp i}=0$ and conservation of the projection $J^3$ of angular
momentum.

The light-front wavefunctions  $\psi_{n/H}(x_i,{\vec k_{\perp i}},
\lambda_i)$ of hadrons are the central elements of QCD
phenomenology, encoding the bound state properties  of hadrons in
terms of their fundamental quark and gluon degrees of freedom at the
amplitude level. It is the probability amplitude that a proton of
momentum $P^+= P^0+P^3$ and transverse momentum $P_\perp$ consists
of $n$ quarks and gluons with helicities $\lambda_i$ and physical
momenta $p^+_i= x_i P^+$ and $p_{\perp i} = x_i P_\perp + k_{\perp
i}$.  The wavefunctions $\{\psi^p_n(x_i, k_{\perp
i},\lambda_i)\},n=3,\dots$ thus describe the proton in an arbitrary
moving frame.  The variables $(x_i, k_{\perp i})$ are internal
relative momentum coordinates.  The fractions $x_i = p^+_i/P^+ =
(p^0_i+p^3_i)/(P^0+P^3)$, $0 <x_i <1$, are the boost-invariant
light-front momentum fractions; $y_i= \log x_i$ is the difference
between the rapidity of the constituent $i$ and the rapidity of the
parent hadron.  The appearance of relative coordinates is connected
to the simplicity of performing Lorentz boosts in the light-front
framework.  This is another major advantage of the light-front
representation.

For example, the eigensolution $\ket{\psi_p}$ of the QCD light-front
Hamiltonian for the proton expanded on the color-singlet $B = 1$, $Q
= 1$ eigenstates $\{\ket{n} \}$ of the free Hamiltonian $
H^{QCD}_{LF}(g = 0).$ This defines the light-front Fock expansion:
\begin{eqnarray}
\ket{ \psi_p(P^+, {\vec P_\perp} )} &=& \sum_{n}\ \prod_{i=1}^{n} {{\rm
d}x_i\, {\rm d}^2
{\vec k_{\perp i}} \over \sqrt{x_i}\, 16\pi^3} \,  \ 16\pi^3  \
\delta\left(1-\sum_{i=1}^{n} x_i\right)\, \delta^{(2)}\left(\sum_{i=1}^{n}
{\vec k_{\perp
i}}\right) \label{a318}
\\
&& \rule{0pt}{4.5ex} \times \psi_{n/H}(x_i,{\vec k_{\perp i}},
\lambda_i) \ket{ n;\, x_i P^+, x_i {\vec P_\perp} + {\vec k_{\perp
i}}, \lambda_i}. \nonumber
\end{eqnarray}
The light-front momentum fractions $x_i = k^+_i/P^+$ and ${\vec k_{\perp i}}$ represent
the relative momentum coordinates of the QCD constituents.  The physical transverse
momenta are ${\vec p_{\perp i}} = x_i {\vec P_\perp} + {\vec k_{\perp i}}.$ The
$\lambda_i$ label the light-front spin projections $S^z$ of the quarks and gluons along
the quantization direction $z$. Each Fock component has the invariant mass squared
\begin{equation}\mathcal{M}^2_n = (\sum^n_{i=1} k_i^\mu)^2
= \sum^n_{i=1}{k^2_{\perp i} + m^2_i\over x_i}.
\end{equation}
The physical gluon polarization vectors $\epsilon^\mu(k,\ \lambda =
\pm 1)$ are specified in light-cone gauge by the conditions $k \cdot
\epsilon = 0,\ \eta \cdot \epsilon = \epsilon^+ = 0.$ The gluonic
quanta which appear in the Fock states thus have physical
polarization $\lambda = \pm 1$ and positive metric. Since each Fock
particle is on its mass shell in a Hamiltonian framework, $k^- =
k^0-k^z= {k^2_\perp + m^2\over k^+}$. The dominant configurations in
the wavefunction are generally those with minimum values of ${\cal
M}^2$.  Note that, except for the case where $m_i=0$ and $k_{\perp
i}=0$, the limit $x_i\rightarrow 0$ is an ultraviolet limit, {\em
i.e.}, it corresponds to particles moving with infinite momentum in
the negative $z$ direction: $k^z_i\rightarrow - k^0_i \rightarrow -
\infty.$

LFWFs have the remarkable property of being independent of the
hadron's four-momentum.  In contrast, in equal-time quantization,  a
Lorentz boost mixes dynamically with the interactions, so that
computing a wavefunction in a new frame at fixed $t$ requires
solving a nonperturbative problem as complicated as the Hamiltonian
eigenvalue problem itself.  The LFWFs are properties of the hadron
itself; they are thus universal and process independent.

The central tool which will be used in these lectures are the
light-front Fock state wavefunctions which encode the bound-state
properties of hadrons in terms of their quark and gluon degrees of
freedom at the amplitude level.  Given these frame-independent
wavefunctions, one can compute a large array of hadronic processes
ranging from the generalized parton distributions measured in deep
inelastic scatterings, hard exclusive reactions, and the weak decays
of hadrons. As I will review below, the quantum fluctuations
contained in the LFWFs  lead to the prediction of novel QCD
phenomena such as color transparency, intrinsic charm, sea quark
asymmetries, and hidden color in nuclear wavefunctions.

Given the light-front wavefunctions $\{\psi_n(x_i,
k_{\perp_i},\lambda_i)\}$ one can compute the electromagnetic and
weak form factors from a simple overlap of light-front
wavefunctions, summed over all Fock states
\cite{Drell:1969km,Brodsky:1980zm}.  Form factors are generally
constructed from hadronic matrix elements of the current $\langle p
\vert j^\mu(0) \vert p + q\rangle,$ where in the interaction picture
we can identify the fully interacting Heisenberg current $J^\mu$
with the free current $j_\mu$ at the spacetime point $x^\mu = 0.$ In
the case of matrix elements of the current $j^+=j^0+j^3$, in a frame
with $q^+=0,$ only diagonal matrix elements in particle number
$n^\prime = n$ are needed.  In contrast, in the equal-time theory
one must also consider off-diagonal matrix elements and fluctuations
due to particle creation and annihilation in the vacuum.  In the
nonrelativistic limit one can make contact with the usual formulae
for form factors in Schr\"odinger many-body theory.

One of the important aspects of fundamental hadron structure is the presence of non-zero
orbital angular momentum in the bound-state wave functions.  The evidence for a ``spin
crisis" in the Ellis-Jaffe sum rule signals a significant orbital contribution in the
proton wave function~\cite{Jaffe:1989jz,Ji:2002qa}.  The Pauli form factor of nucleons is
computed from the overlap of LFWFs differing by one unit of orbital angular momentum
$\Delta L_z= \pm 1$.  Thus the fact that the anomalous moment of the proton is non-zero
requires nonzero orbital angular momentum in the proton
wavefunction~\cite{Brodsky:1980zm}.  In the light-front method, orbital angular momentum
is treated explicitly; it includes the orbital contributions induced by relativistic
effects, such as the spin-orbit effects normally associated with the conventional Dirac
spinors.  Angular momentum conservation for each Fock state implies
\begin{equation}
J^z= \sum_i^{n} S^z_i + \sum_i^{n-1} L^z_i
\end{equation}
where $L^z_i$ is one of the $n-1$ relative orbital angular momenta.

The quark and gluon probability distributions of a hadron are
constructed from integrals over the absolute squares $\vert \psi_n
\vert^2 $, summed over $n.$ In the far off-shell domain of large
parton virtuality, one can use perturbative QCD or conformal
arguments to derive the asymptotic fall-off of the Fock amplitudes,
which then in turn leads to the QCD evolution equations for
distribution amplitudes and structure functions. More generally, one
can prove factorization theorems for exclusive and inclusive
reactions which separate the hard and soft momentum transfer
regimes, thus obtaining rigorous predictions for the leading power
behavior contributions to large momentum transfer cross sections.
One can also compute the far off-shell amplitudes within the
light-front wavefunctions where heavy quark pairs appear in the Fock
states.  Such states persist over a time $\tau \simeq P^+/{\cal
M}^2$ until they are materialized in the hadron collisions.  As I
shall discuss, this leads to a number of novel effects in the
hadroproduction of heavy quark hadronic states.

\section{Light-Front Statistical Physics}

As shown by Raufeisen and myself~\cite{Raufeisen:2004dg}, one can
construct a ``light-front density matrix" from the complete set of
light-front wavefunctions which is a Lorentz scalar. This form can
be used at finite temperature to give a boost invariant formulation
of thermodynamics.  At zero temperature the light-front density
matrix is directly connected to the Green's function for quark
propagation in the hadron as well as deeply virtual Compton
scattering. One can also define a light-front partition function
$Z_{LF}$ as an outer product of light-front wavefunctions. The
deeply virtual Compton amplitude and generalized parton
distributions can then be computed as the trace $Tr[Z_{LF}
\mathcal{O}],$ where $\mathcal{O}$ is the appropriate local
operator~\cite{Raufeisen:2004dg}. This partition function formalism
can be extended to multi-hadronic systems and systems in statistical
equilibrium to provide a Lorentz-invariant description of
relativistic thermodynamics~\cite{Raufeisen:2004dg}.

\section{AdS/CFT and Hadron Phenomenology}

Maldacena~\cite{Maldacena:1997re} has shown that there is a
remarkable correspondence between large $N_C$ supergravity theory in
a higher dimensional  anti-de Sitter space and supersymmetric QCD in
4-dimensional space-time.  String/gauge duality provides a framework
for predicting QCD phenomena based on the conformal properties of
the AdS/CFT correspondence. For example, Polchinski and
Strassler~\cite{Polchinski:2001tt} have shown that the power-law
fall-off of hard exclusive hadron-hadron scattering amplitudes at
large momentum transfer can be derived without the use of
perturbation theory by using the scaling properties of the hadronic
interpolating fields in the large-$r$ region of AdS space.  Thus one
can use the Maldacena correspondence to compute the leading
power-law falloff of exclusive processes such as high-energy
fixed-angle scattering of gluonium-gluonium scattering in
supersymmetric QCD.   The resulting predictions for hadron physics
effectively
coincide~\cite{Polchinski:2001tt,Brower:2002er,Andreev:2002aw} with
QCD dimensional counting
rules~\cite{Brodsky:1973kr,Matveev:1973ra,Brodsky:1974vy,Brodsky:2002st}.
Polchinski and Strassler~\cite{Polchinski:2001tt} have also derived
counting rules for deep inelastic structure functions at $x \to 1$
in agreement with perturbative QCD predictions~\cite{Brodsky:1994kg}
as well as Bloom-Gilman exclusive-inclusive duality.  An interesting
point is that the hard scattering amplitudes which are normally or
order $\alpha_s^p$ in PQCD appear as order $\alpha_s^{p/2}$ in the
supergravity predictions.  This can be understood as an all-orders
resummation of the effective
potential~\cite{Maldacena:1997re,Rey:1998ik}.  The near-conformal
scaling properties of light-front wavefunctions thus lead to a
number of important predictions for QCD which are normally discussed
in the context of perturbation theory.

De Teramond and I~\cite{Brodsky:2003px} have shown how one can use
the scaling properties of the hadronic interpolating operator in the
extended AdS/CFT space-time theory to determine the form of QCD
wavefunctions at large transverse momentum $k^2_\perp \to \infty$
and at $x \to 1$~\cite{Brodsky:2003px}.  The angular momentum
dependence of the light-front wavefunctions also follow from the
conformal properties of the AdS/CFT correspondence.  The scaling and
conformal properties of the correspondence leads to a hard component
of the light-front Fock state wavefunctions of the form:
\begin{eqnarray}
 \psi_{n/h} (x_i, \vec k_{\perp i} , \lambda_i, l_{z i})
&\sim& \frac{(g_s~N_C)^{\frac{1}{2} (n-1)}}{\sqrt {N_C}}
 ~\prod_{i =1}^{n - 1} (k_{i \perp}^\pm)^{\vert l_{z i}\vert}\\[1ex]
&&\times \left[\frac{ \Lambda_o}{ {M}^2 - \sum _i\frac{\vec k_{\perp i}^2 +
m_i^2}{x_i} +
\Lambda_o^2}  \right] ^{n +\sum_i \vert l_{z i} \vert -1}\ ,\nonumber
\label{eq:lfwfR}
\end{eqnarray}
where $g_s$ is the string scale and $\Lambda_o$ represents the basic QCD mass scale.  The
scaling predictions agree with the perturbative QCD analysis given in the
references~\cite{Ji:2003fw}, but the AdS/CFT analysis is performed at strong coupling
without the use of perturbation theory.  The form of these near-conformal wavefunctions
can be used as an initial ansatz for a variational treatment of the light-front QCD
Hamiltonian.  The same ansatz leads to predictions for the hadron spectrum,which I will
discuss in the conclusions.

\section{Light-Front Wavefunctions and Hadron Phenomenology}

Even though QCD was motivated by the successes of the parton model,
QCD predicts many new features which go well beyond the simple
three-quark description of the proton.  Since the number of Fock
components cannot be limited in relativity and quantum mechanics,
the nonperturbative wavefunction of a proton contains gluons and sea
quarks, including heavy quarks at any resolution scale. Thus there
is no scale $Q_0$ in deep inelastic lepton-proton scattering where
the proton can be approximated by its valence quarks. The
nonperturbative Fock state wavefunctions contain intrinsic gluons,
strange quarks, charm quarks, etc., at any scale.  The internal QCD
interactions lead to asymmetries such as $ s(x) \ne \bar s(x)$,
$\bar u(x) \ne \bar d(x)$ and intrinsic charm and bottom
distributions at large $x$ since this minimizes the invariant mass
and off-shellness of the higher Fock state.  As discussed above, the
Fock state expansion for nuclei contains hidden color states which
cannot be classified in terms of nucleonic degrees of freedom.
However, some leading-twist phenomena such as the diffractive
component of deep inelastic scattering, single-spin asymmetries,
nuclear shadowing and antishadowing cannot be computed from the
LFWFs of hadrons in isolation.

\subsection{The Strange Quark Asymmetry}

In the simplest treatment of deep inelastic scattering, nonvalence
quarks are produced via gluon splitting and DGLAP evolution.
However, in the full theory, heavy quarks are multiply connected to
the valence quarks~\cite{Brodsky:1980pb}. Although the strange and
antistrange distributions in the nucleon are identical when they
derive from gluon-splitting $g \to s \bar s$, this is not the case
when the strange quarks are part of the intrinsic structure of the
nucleon -- the multiple interactions of the sea quarks produce an
asymmetry of the strange and anti-strange distributions in the
nucleon due to their different interactions with the other quark
constituents.  A QED analogy is the distribution of $\tau^+$ and
$\tau^-$ in a higher Fock state of muonium $\mu^+ e^-.$  The
$\tau^-$ is attracted to the higher momentum $\mu^+$ thus
asymmetrically distorting its momentum distribution. Similar effects
will  happen in QCD. If we use the diquark model $\ket p  \sim
\ket{u_{3_c} (ud)_{\bar 3_C}},$ then the $Q_{3_C}$ in the $\ket{u
(ud) Q \bar Q }$ Fock state will be attracted to the heavy diquark
and thus have higher rapidity than the $\bar Q$. An alternative
model is the $\ket{K \Lambda}$ fluctuation model for the $\ket{uud s
\bar s}$ Fock state of the proton~\cite{Brodsky:1996hc}. The $s$
quark tends to have higher $x$.

\begin{figure}[htb]
\begin{center}
\includegraphics[width=4in]{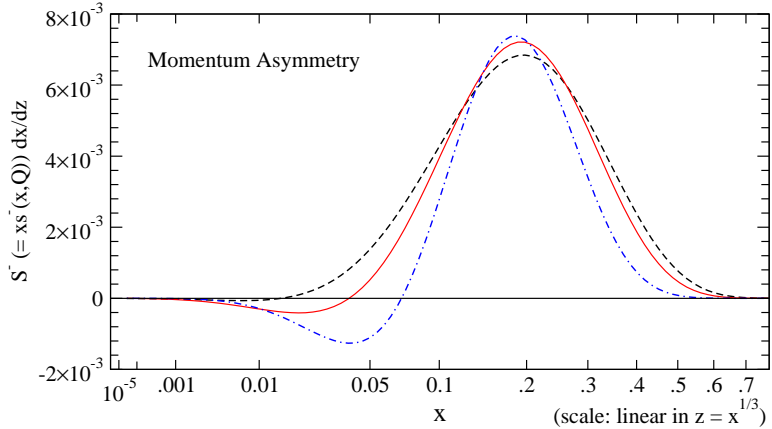}
\end{center}
\caption{Representative results of the CTEQ strangeness
asymmetry analysis.\label{fig:sasym}}
\end{figure}

Empirical evidence also continues to accumulate that the
strange-antistrange quark distributions are not symmetric in the
proton~\cite{Brodsky:1996hc,Kretzer:2004bg,Portheault:2004xy}. The
experimentally observed asymmetry appears to be small but positive:
$\int dx x [s(x)- \bar s(x)] > 0.$ The results of a recent CTEQ
global data analysis \cite{Olness:2003wz} of neutrino-induced dimuon
data are shown in Fig.~\ref{fig:sasym}.
The fit is  constrained so that the number of $s$ and $\bar s$
quarks in the nucleon are equal. The shape of the strangeness
asymmetry is consistent with the $\Lambda K$ fluctuation
model~\cite{Brodsky:1996hc}. Kretzner~\cite{Kretzer:2004bg} has
noted that a significant part of the NuTeV anomaly could be due to
this asymmetry, The $\bar s(x)-s(x)$ asymmetry can be studied in
detail in $p \bar p$ collisions by searching for antisymmetric
forward-backward strange quark distributions in the $\bar p-p$ CM
frame.

\subsection{Intrinsic Heavy Quarks}

The probability for Fock states of a light hadron such as the proton
to have an extra heavy quark pair decreases as $1/m^2_Q$ in
non-Abelian gauge theory~\cite{Franz:2000ee,Brodsky:1984nx}.  The
relevant matrix element is the cube of the QCD field strength
$G^3_{\mu \nu}.$  This is in contrast to abelian gauge theory where
the relevant operator is $F^4_{\mu \nu}$ and the probability of
intrinsic heavy leptons in QED bound state is suppressed as
$1/m^4_\ell.$  The intrinsic Fock state probability is maximized at
minimal off-shellness.  It is useful to define the transverse mass
$m_{\perp i}= \sqrt{k^2_{\perp i} + m^2_i}.$ The maximum probability
then occurs at $x_i = { m^i_\perp /\sum^n_{j = 1} m^j_\perp}$; {\em
i.e.}, when the constituents have minimal invariant mass and equal
rapidity. Thus the heaviest constituents have the highest momentum
fractions and the highest $x_i$. Intrinsic charm thus predicts that
the charm structure function has support at large $x_{bj}$ in excess
of DGLAP extrapolations~\cite{Brodsky:1980pb}; this is in agreement
with the EMC measurements~\cite{Harris:1995jx}.

Intrinsic charm can also explain the $J/\psi \to \rho \pi$
puzzle~\cite{Brodsky:1997fj}. It also affects the extraction of
suppressed CKM matrix elements in $B$ decays~\cite{Brodsky:2001yt}.

\subsection{Diffractive Dissociation and Intrinsic Heavy Quark Production}

\begin{figure}[ht]
\begin{center}
\includegraphics[height=2.5in]{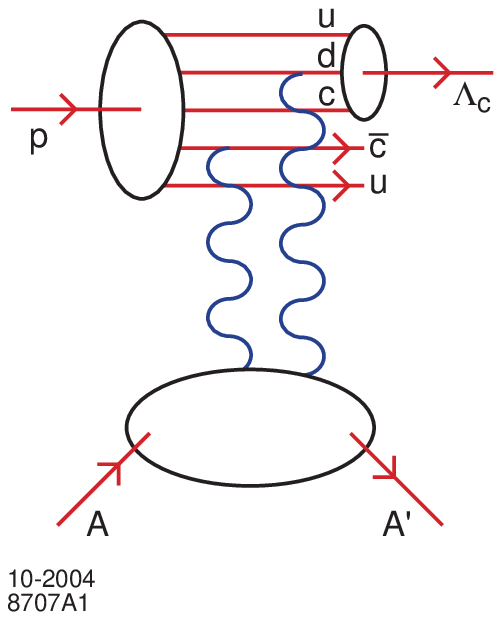}
\end{center}
\caption[*]{\baselineskip 13pt Production of forward heavy baryons
by diffractive dissociation. \label{DD}}
\end{figure}

Diffractive dissociation is particularly relevant to the production
of leading heavy quark states.  The projectile proton can be
decomposed as a sum over all of its Fock state components.  The
diffractive dissociation of the intrinsic charm $|uud c \bar c>$
Fock state of the proton on a nucleus can produce a leading heavy
quarkonium state at high $x_F = x_c + x_{\bar c}~$ in $p A \to
J/\psi X A^\prime$ since the $c$ and $\bar c$ can readily coalesce
into the charmonium state.  Since the constituents of a given
intrinsic heavy-quark Fock state tend to have the same rapidity,
coalescence of multiple partons from the projectile Fock state into
charmed hadrons and mesons is also favored.  For example, as
illustrated in  Fig. \ref{DD}, one can produce leading $\Lambda_c$
at high $x_F$ and low $p_T$ from the coalescence of the $u d c$
constituents of the projectile IC Fock state.  A similar coalescence
mechanism was used in atomic physics to produce relativistic
antihydrogen in $\bar p A$ collisions~\cite{Munger:1993kq}. This
phenomena is important not only for understanding heavy-hadron
phenomenology, but also for understanding the sources of neutrinos
in astrophysics experiments~\cite{Halzen:2004bn}.

\vspace{0.3cm}

The charmonium state will be produced at small transverse momentum
and high $x_F$  with a characteristic $A^{2/3}$ nuclear dependence.
This forward contribution is in addition to the $A^1$ contribution
derived from the usual perturbative QCD fusion contribution at small
$x_F.$   Because of these two components, the cross section violates
perturbative QCD factorization for hard inclusive
reactions~\cite{Hoyer:1990us}.  This is consistent with the observed
two-component cross section for charmonium production observed by
the NA3 collaboration at CERN~\cite{Badier:1981ci}.

The diffractive dissociation of the intrinsic charm Fock state leads
to leading charm hadron production and fast charmonium production in
agreement with measurements~\cite{Anjos:2001jr}.    Intrinsic charm
can also explain the $J/\psi \to \rho \pi$
puzzle~\cite{Brodsky:1997fj},  and it affects the extraction of
suppressed CKM matrix elements in $B$ decays~\cite{Brodsky:2001yt}.
Intrinsic charm can also enhance the production probability of Higgs
bosons at hadron colliders from processes such as $g c \to H c.$ It
is thus critical for new experiments (HERMES, HERA, COMPASS) to
definitively establish the phenomenology of the charm structure
function at large $x_{bj}.$

The production cross section for the double-charm $\Xi_{cc}^+$
baryon~\cite{Ocherashvili:2004hi} and the production of $J/\psi$
pairs appears to be consistent with the diffractive dissociation and
coalescence of double IC Fock states~\cite{BGK,Vogt:1995tf}.  It is
unlikely that the appearance of two heavy quarks at high $x_F$ could
be explained by the ``color drag model" used in PYTHIA
simulations~\cite{Andersson:1983ia} in which the heavy quarks are
accelerated from low to high $x$ by the fast valence quarks. These
observations provide compelling evidence for the diffractive
dissociation of complex off-shell Fock states of the projectile and
contradict the traditional view that sea quarks and gluons are
always produced perturbatively via DGLAP evolution. It is also
conceivable that the observations~\cite{Bari:1991ty} of $\Lambda_b$
at high $x_F$ at the ISR in high energy $p p$  collisions could be
due to the diffractive dissociation and coalescence of the
``intrinsic bottom" $|uud b \bar b>$ Fock states of the proton.

\subsection{ Color transparency}  The small transverse size
fluctuations of a hadron wavefunction with a small color dipole
moment will have minimal interactions in a
nucleus~\cite{Bertsch:1981py,Brodsky:1988xz}.

This has been verified in the case of diffractive dissociation  of a
high energy pion into dijets $\pi A \to q \bar q A^\prime$ in which
the nucleus is left in its ground state~\cite{Ashery:2002jx}. As
discussed in the next subsection, when the hadronic jets have
balancing but high transverse momentum, one studies the small size
fluctuation of the incident pion.  The diffractive dissociation
cross section is found to be proportional to $A^2$ in agreement with
the color transparency prediction.

Color transparency has also been observed in diffractive
electroproduction of $\rho$ mesons \cite{Borisov:2002rd} and in
quasi-elastic $p A \to p p (A-1)$ scattering~\cite{Aclander:2004zm}
where only the small size fluctuations of the hadron wavefunction
enters the hard exclusive scattering amplitude.  In the latter case
an anomaly occurs at $\sqrt s \simeq 5 $ GeV, most likely signaling
a resonance effect at the charm threshold~\cite{Brodsky:1987xw}.

\subsection{Diffraction Dissociation as a Tool to Resolve Hadron Substructure}

Diffractive multi-jet production in heavy nuclei provides a novel
way to measure the shape of light-front Fock state wave functions
and test color transparency~\cite{Brodsky:1988xz}.  For example,
consider the reaction~\cite{Bertsch:1981py,Frankfurt:1999tq} $\pi A
\rightarrow {\rm Jet}_1 + {\rm Jet}_2 + A^\prime$ at high energy
where the nucleus $A^\prime$ is left intact in its ground state.
The transverse momenta of the jets balance so that $ \vec k_{\perp
i} + \vec k_{\perp 2} = \vec q_\perp < {R^{-1}}_A \ . $ The
light-front longitudinal momentum fractions also need to add to
$x_1+x_2 \sim 1.$   Diffractive dissociation on a nucleus also
requires that the energy of the beam has to be sufficiently large
such that the momentum transfer to the nucleus $\Delta p_L = {\Delta
M^2\over 2 E_{lab}}$ is smaller than the inverse nuclear size $R_A.$
The process can then occur coherently in the nucleus.

Because of color transparency, the valence wave function of the pion
with small impact separation will penetrate the nucleus with minimal
interactions, diffracting into jet pairs~\cite{Bertsch:1981py}.  The
$x_1=x$, $x_2=1-x$ dependence of the di-jet distributions will thus
reflect the shape of the pion valence light-front wave function in
$x$; similarly, the $\vec k_{\perp 1}- \vec k_{\perp 2}$ relative
transverse momenta of the jets gives key information on the second
transverse momentum derivative of the underlying shape of the
valence pion wavefunction~\cite{Frankfurt:1999tq,Nikolaev:2000sh}.
The diffractive nuclear amplitude extrapolated to $t = 0$ should be
linear in nuclear number $A$ if color transparency is correct.  The
integrated diffractive rate will then scale as $A^2/R^2_A \sim
A^{4/3}.$ This is in fact what has been observed by the E791
collaboration at FermiLab for 500 GeV incident pions on nuclear
targets~\cite{Aitala:2000hc}.  The measured momentum fraction
distribution of the jets is found to be approximately consistent
with the shape of the pion asymptotic distribution amplitude,
$\phi^{\rm asympt}_\pi (x) = \sqrt 3 f_\pi
x(1-x)$~\cite{Aitala:2000hb}. Data from CLEO~\cite{Gronberg:1998fj}
for the $\gamma \gamma^* \rightarrow \pi^0$ transition form factor
also favor a form for the pion distribution amplitude close to the
asymptotic solution to its perturbative QCD evolution
equation~\cite{Lepage:1979zb,Efremov:1978rn,Lepage:1980fj}.

Color transparency, as evidenced by the Fermilab measurements of
diffractive dijet production, implies that a pion can interact
coherently throughout a nucleus with minimal absorption, in dramatic
contrast to traditional Glauber theory based on a fixed $\sigma_{\pi
n}$ cross section.  Color transparency gives direct validation of
the gauge interactions of QCD.

\subsection{Diffractive Dissociation and Hidden Color in Nuclear Wavefunctions}

The concept of high energy diffractive dissociation can be
generalized to provide a tool to materialize the individual Fock
states of a hadron, nucleus or photon. For example, the diffractive
dissociation of a high energy proton on a nucleus $p A \to X
A^\prime$ where the diffractive system is three jets $X= q q q$ can
be used to determine the valence light-front wavefunction of the
proton.

In the case of a deuteron projectile, one can study diffractive
processes such as $d A \to p n A^\prime$ or $d A \to \pi^- p p$ to
measure the mesonic Fock state of a nuclear wavefunction.  At small
hadron transverse momentum, diffractive dissociation of the deuteron
should be controlled by conventional nuclear interactions; however
at large relative $k_T$, the diffractive system should be sensitive
to the hidden color components of the deuteron wavefunction.
The theory of hidden color is reviewed below.

\section{DLCQ Solutions}

The entire spectrum of hadrons and nuclei and their scattering
states is given by the set of eigenstates of the light-front
Hamiltonian $H_{LC}$ for QCD. In principle it is possible to compute
the light-front wavefunctions by diagonalizing the QCD light-front
Hamiltonian on the free Hamiltonian basis.  In the case of QCD in
one space and one time dimensions, the application of discretized
light-front quantization (DLCQ) \cite{Pauli:1985pv} provides
complete solutions of the theory, including the entire spectrum of
mesons, baryons, and nuclei, and their wavefunctions.  In the DLCQ
method, one uses periodic boundary conditions in $x^-$ and $b_\perp$
to discretize the light-front momentum space. One then diagonalizes
the light-front Hamiltonian for QCD on a discretized Fock state
basis. The DLCQ solutions can be obtained for arbitrary parameters
including the number of flavors and colors and quark masses. Exact
solutions are known  for $QCD(1+1)$ at $N_C \to \infty$  by 't
Hooft~\cite{'tHooft:1974hx}. The one-space one-time theory can be
solved numerically to any precision at finite $N_C$ for any coupling
strength and number of quark flavors using discretized light-front
quantization
(DLCQ)~\cite{Pauli:1985pv,Hornbostel:1988fb,burkardt89,Brodsky:1997de}.
One can use DLCQ to calculate the entire spectrum of virtually any
1+1 theory, its discrete bound states as well as the scattering
continuum. The main emphasis of the DLCQ method applied to QCD is
the determination of the wavefunctions of the hadrons from first
principles.

A large number of studies have been performed of model field
theories in the LF framework.  This approach has been remarkably
successful in a range of toy models in 1+1 dimensions: Yukawa theory
\cite{Pauli:1985ps}, the Schwinger model (for both massless and
massive fermions)
\cite{Eller:1986nt,McCartor:1991gn,McCartor:1994im}, $\phi^4$ theory
\cite{Harindranath:1987db,Harindranath:1987ex}, QCD with various
types of matter
\cite{Burkardt:1989wy,Hornbostel:1988fb,Demeterfi:1993rs,Dalley:1992yy},
and the sine-Gordon model \cite{Burkardt:1992sz}.  It has also been
applied with promising results to theories in 3+1 dimensions, in
particular QED \cite{Krautgartner:1991xz,Kaluza:1991kx} and Yukawa
theory \cite{Brodsky:2002tp} in a truncated basis.  In all cases
agreement was found between the LC calculations and results obtained
by more conventional approaches, for example, lattice gauge theory.

The extension of this program to physical theories in 3+1 dimensions
is a formidable computational task because of the much larger number
of degrees of freedom; however, progress is being made.  Analyses of
the spectrum and light-front wavefunctions of positronium in
QED$_{3+1}$ are given in Ref. \cite{Krautgartner:1991xz}.

\subsection{A DLCQ example: QCD$_{1+1}$ with Fundamental Matter }

This theory was originally considered by 't Hooft in the limit of
large $N_c$ \cite{'tHooft:1974hx}.  Later Burkardt
\cite{Burkardt:1989wy}, and Hornbostel, {\em et al.}
\cite{Hornbostel:1988fb}, gave essentially complete numerical
solutions of the theory for finite $N_c$, obtaining the spectra of
baryons, mesons, and nucleons and their wavefunctions.  The DLCQ
results are consistent with the few other calculations available for
comparison, and are generally much more efficiently obtained.  In
particular, the mass of the lowest meson agrees to within numerical
accuracy with lattice Hamiltonian results \cite{Hamer:1981yq}.  For
$N_c=4$ this mass is close to that obtained by 't Hooft in the
$N_c\rightarrow\infty$ limit \cite{'tHooft:1974hx}.  Finally, the
ratio of baryon to meson mass as a function of $N_c$ agrees with the
strong-coupling results of Ref. \cite{Date:1986xe}.

\begin{figure}[htb]
\begin{center}
\includegraphics[width=4in]{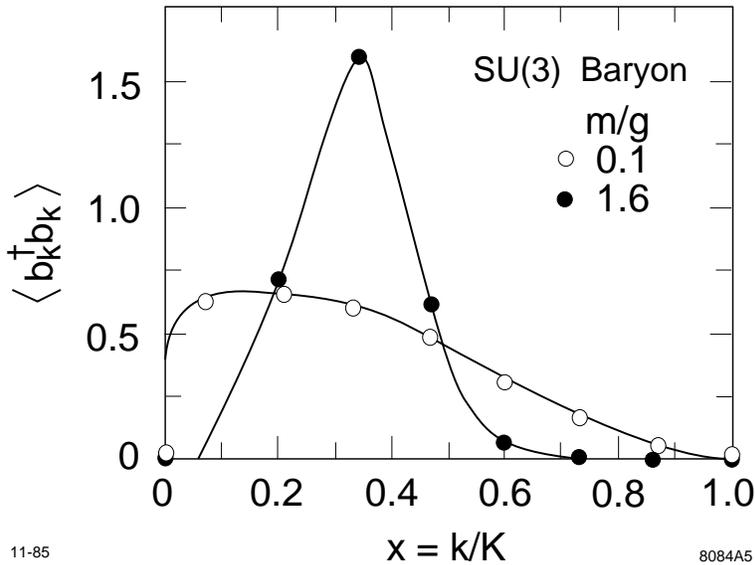}
\end{center}
\caption{Valence contribution to the baryon structure function in
QCD$_{1+1}$, as a function of the light-front longitudinal momentum
fraction.  The gauge group is SU(3), $m$ is the quark mass, and $g$
is the gauge coupling. (From Ref. [\cite{Hornbostel:1988fb}].) }
\label{fig1}
\end{figure}

\begin{figure}[htb]
\begin{center}
\includegraphics[width=4.75in,height=2.5in]{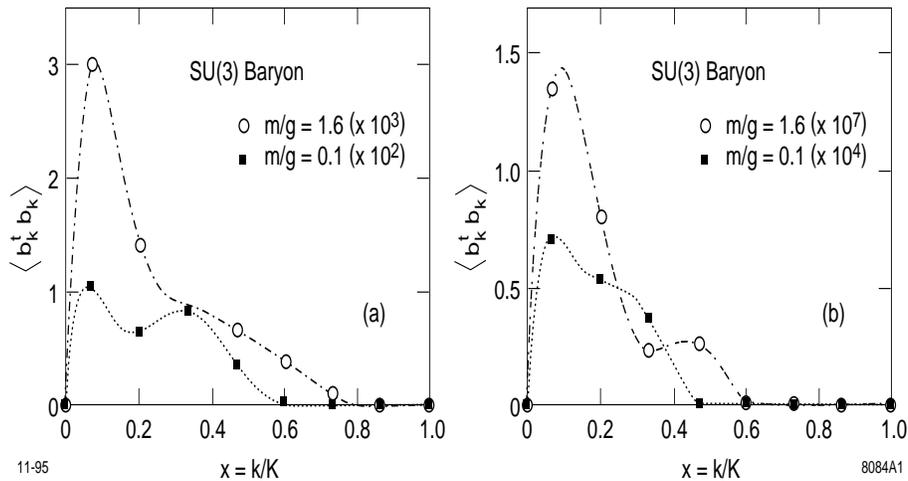}
\end{center} \caption{Contributions to the baryon structure function from
higher Fock components: (a) valence plus one additional $q\bar{q}$
pair; (b) valence plus two additional $q\bar{q}$ pairs.  (From Ref.
[\cite{Hornbostel:1988fb}].) } \label{fig2}
\end{figure}

In addition to the spectrum, one obtains the wavefunctions. These
allow direct computation of, {\em e.g.}, structure functions.  As an
example, Fig.~\ref{fig1} shows the valence contribution to the
structure function for an SU(3) baryon, for two values of the
dimensionless coupling $m/g$.  As expected, for weak coupling the
distribution is peaked near $x=1/3$, reflecting that the baryon
momentum is shared essentially equally among its constituents.  For
comparison, the contributions from Fock states with one and two
additional $q\bar{q}$ pairs are shown in Fig.~\ref{fig2}. Note that
the amplitudes for these higher Fock components are quite small
relative to the valence configuration.  The lightest hadrons are
nearly always dominated by the valence Fock state in these
super-renormalizable models; higher Fock wavefunctions are typically
suppressed by factors of 100 or more.  Thus the light-front quarks
are much more like constituent quarks in these theories than
equal-time quarks would be. As discussed above, in an equal-time
formulation even the vacuum state would be an infinite superposition
of Fock states.  Identifying constituents in this case, three of
which could account for most of the structure of a baryon, would be
quite difficult.

\section{Light-Front Wavefunctions and Hadron Observables}

Light-front Fock state wavefunctions $\psi_{n/H}(x_i,\vec k_{\perp i},\lambda_i)$ play an
essential role in QCD  phenomenology, generalizing Schr\"odinger wavefunctions
$\psi_H(\vec k)$ of atomic physics to relativistic quantum field theory. Given the
$\psi^{(\Lambda)}_{n/H},$ one can construct any spacelike electromagnetic, electroweak,
or gravitational form factor or local operator product matrix element of a composite or
elementary system from the diagonal overlap of the LFWFs~\cite{Brodsky:1980zm}. Exclusive
semi-leptonic $B$-decay amplitudes involving timelike currents such as $B\rightarrow A
\ell \bar{\nu}$ can also be evaluated exactly in the light-front
formalism~\cite{Brodsky:1998hn}.  In this case, the timelike decay matrix elements
require the computation of both the diagonal matrix element $n \rightarrow n$ where
parton number is conserved and the off-diagonal $n+1\rightarrow n-1$ convolution such
that the current operator annihilates a $q{\bar{q'}}$ pair in the initial $B$
wavefunction.  This term is a consequence of the fact that the time-like decay $q^2 =
(p_\ell + p_{\bar{\nu}} )^2 > 0$ requires a positive light-front momentum fraction $q^+ >
0$. Conversely for space-like currents, one can choose $q^+=0$, as in the Drell-Yan-West
representation of the space-like electromagnetic form factors. The light-front Fock
representation thus provides an exact formulation of current matrix elements of local
operators.  In contrast, in equal-time Hamiltonian theory, one must evaluate connected
time-ordered diagrams where the gauge particle or graviton couples to particles
associated with vacuum fluctuations.  Thus even if one knows the equal-time wavefunction
for the initial and final hadron, one cannot determine the current matrix elements.  In
the case of the covariant Bethe-Salpeter formalism, the evaluation of the matrix element
of the current requires the calculation of an infinite number of irreducible diagram
contributions.

One can also prove directly from the LFWF overlap representation
that the anomalous gravitomagnetic moment $B(0)$ vanishes for any
composite system~\cite{Brodsky:2000ii}.  This property follows
directly from the Lorentz boost properties of the light-front Fock
representation and holds separately for each Fock state component.

Given the light-front wave functions, one can define
positive-definite probability distributions, such as the quark and
gluon distributions $q(x,Q)$, $g(x,Q)$ which enter deep inelastic
scattering and other hard inclusive reactions. This include all
spin-dependent distributions such as quark transversity. The
resulting distributions obey DGLAP evolution; the moments defined as
the matrix elements of the operator product expansion have the
correct anomalous dimensions. In addition one can compute the
unintegrated distributions in $x$ and $k_\perp$ which underlie the
generalized parton distributions for nonzero skewness. 
example, the polarized quark distributions at resolution $\Lambda$
correspond to
\begin{eqnarray}
q_{\lambda_q/\Lambda_p}(x, \Lambda) &=&  \sum_{n,q_a}
\int\prod^n_{j=1} dx_j d^2 k_{\perp j}\sum_{\lambda_i} \vert
\psi^{(\Lambda)}_{n/H}(x_i,\vec k_{\perp i},\lambda_i)\vert^2
 \\
&& \times\ \delta\left(1- \sum^n_i x_i\right) \delta^{(2)}
\left(\sum^n_i \vec k_{\perp i}\right) \delta(x - x_q)\nonumber \\
&& \times\  \delta_{\lambda_a, \lambda_q} \Theta(\Lambda^2 -
\mathcal{M}^2_n)\ ,\nonumber
\end{eqnarray}
where the sum is over all quarks $q_a$ which match the quantum
numbers, light-front momentum fraction $x,$ and helicity of the
struck quark.

Diehl, Hwang, and I~\cite{Brodsky:2000xy} have shown how to represent virtual Compton
scattering $\gamma^* p \to \gamma p$ at large initial photon virtuality $Q^2$ and small
momentum transfer squared $t$ in handbag approximation in terms of the light-front
wavefunctions of the target proton. Thus the generalized parton distributions which enter
virtual Compton scattering and the two-photon exchange contribution to lepton-proton
scattering are given by overlaps of the LFWFS with $n = n^\prime$ and $n- n^\prime = \pm
2$. One can  verify that the skewed parton distributions $H(x,\zeta,t)$ and
$E(x,\zeta,t)$ which appear in deeply virtual Compton scattering are the integrands of
the Dirac and Pauli form factors $F_1(t)$ and $F_2(t)$ and the gravitational form factors
$A_{q}(t)$ and $B_{q}(t)$ for each quark and anti-quark constituent. We have given an
explicit illustration of the general formalism for the case of deeply virtual Compton
scattering on the quantum fluctuations of a fermion in quantum electrodynamics at one
loop. The absolute square of the LFWFS define the unintegrated parton distributions. The
integrals of the unintegrated parton distributions over transverse momentum at  zero
skewness provide the helicity and transversity distributions measurable in polarized deep
inelastic experiments \cite{Lepage:1980fj}.

\begin{figure}
\begin{center}
\includegraphics{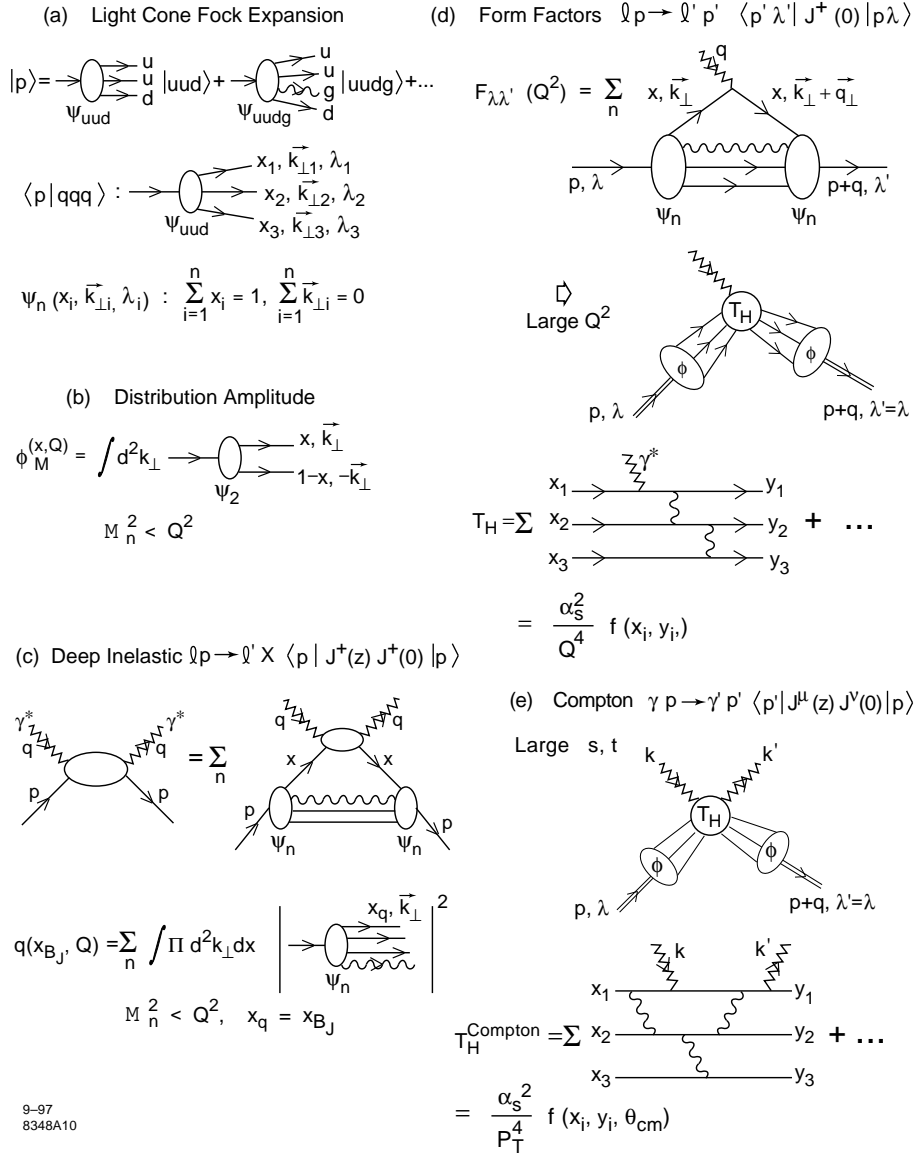}
\end{center}
\caption[*]{ Representation of QCD hadronic processes in the
light-front Fock expansion. (a) The valence $uud$  and higher Fock
$uudg$ contributions to the light-front Fock expansion for the
proton. (b) The distribution amplitude $\phi(x,Q)$ of a meson
expressed as an integral over its valence light-front wavefunction
restricted to $q \bar q$ invariant mass less than $Q$.  (c)
Representation of deep inelastic scattering and the quark
distributions $q(x,Q)$  as probabilistic measures of the light-front
Fock wavefunctions. The sum is over the Fock states with invariant
mass less than $Q$. (d) Exact representation of spacelike form
factors of the proton in the light-front Fock basis.  The sum is
over all Fock components. At large momentum transfer the
leading-twist contribution factorizes as the product of the hard
scattering amplitude $T_H$ for the scattering of the valence quarks
collinear with the initial to final direction convoluted with the
proton distribution amplitude.  (e) Leading-twist factorization of
the Compton amplitude at large momentum transfer. \label{Fig:repc1}}
\end{figure}

\begin{figure}[htbp]
\includegraphics{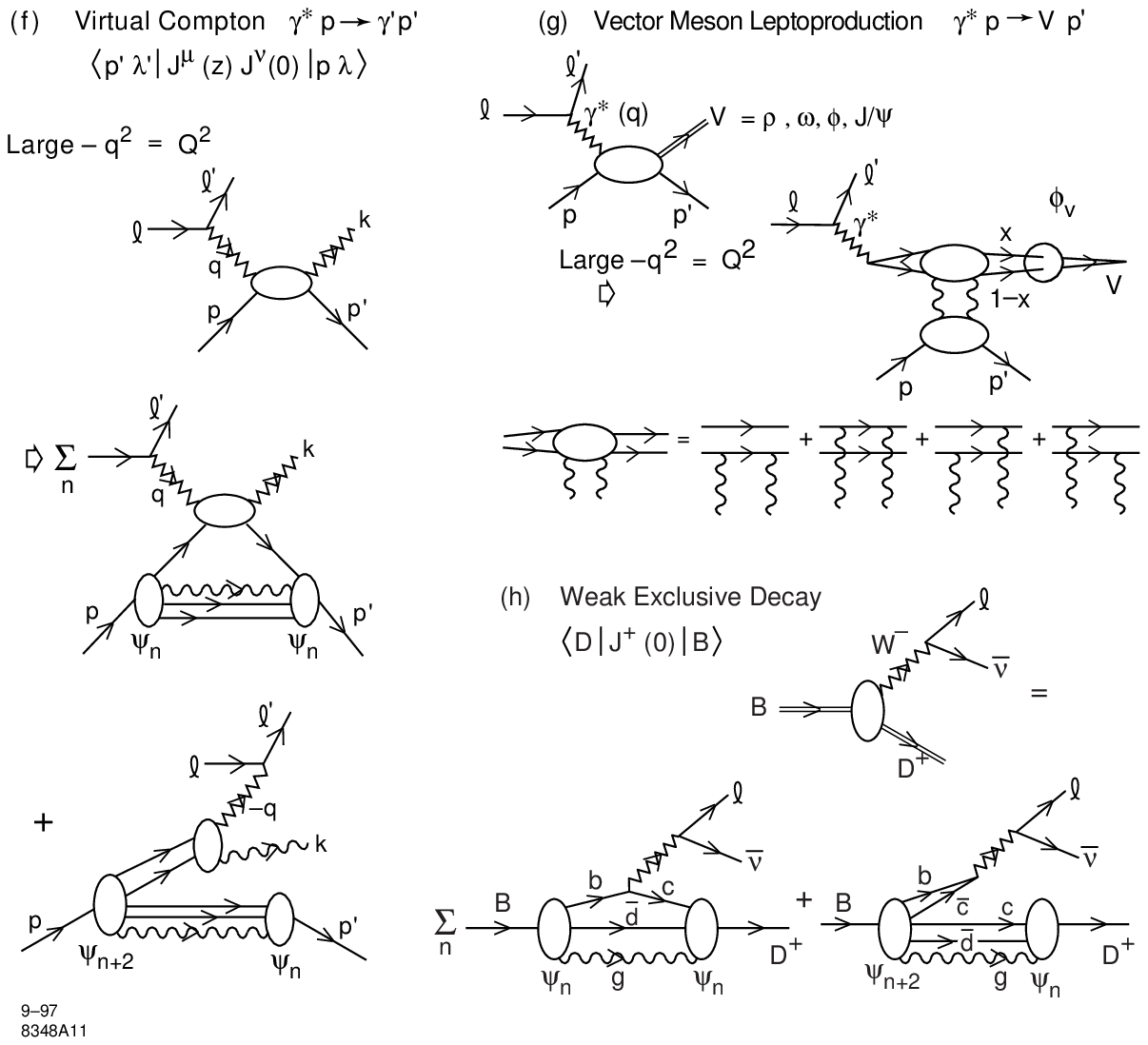}
\caption{ (f) Representation of deeply virtual Compton scattering in
the light-front Fock expansion at leading twist. Both diagonal $n
\to n$ and off-diagonal $n+2 \to n$ contributions are required.  (g)
Diffractive vector meson production at large photon virtuality $Q^2$
and longitudinal polarization.  The high energy behavior involves
two gluons in the $t$ channel coupling to the compact color dipole
structure of the upper vertex.  The bound-state structure of the
vector meson enters through its distribution amplitude.  (h) Exact
representation of the weak semileptonic decays of heavy hadrons in
the light-front Fock expansion.   Both diagonal $n  \to n$ and
off-diagonal pair annihilation $n+2 \to n$ contributions are
required. \label{Fig:repc2}}
\end{figure}

The relationship of QCD processes to the hadron LFWFs is illustrated in
Figs.~\ref{Fig:repc1} and \ref{Fig:repc2}.
Other applications include two-photon exclusive reactions, and
diffractive dissociation into jets.  The universal light-front
wave functions and distribution amplitudes control hard exclusive
processes such as form factors, deeply virtual Compton scattering,
high momentum transfer photoproduction, and two-photon processes.

\vspace{.5cm}

\vspace{.5cm}

Hadronization phenomena such as the coalescence mechanism for leading heavy hadron
production can also be computed from LFWF overlaps. Diffractive jet production provides
another phenomenological window into the structure of LFWFs.  However, as shown
recently~\cite{Brodsky:2002ue} and discussed below, some leading-twist phenomena such as
the diffractive component of deep inelastic scattering, single spin asymmetries, nuclear
shadowing and antishadowing cannot be computed from the LFWFs of hadrons in isolation.

Given the LFWFs, one can also compute the hadronic distribution
amplitudes \break $\phi_H(x_i,Q)$  which control hard exclusive
processes as an integral over the transverse momenta of the valence
Fock state LFWFs~\cite{Lepage:1980fj}. The hadron distribution
amplitudes are obtained by integrating the $n-$parton valence
light-front wavefunctions: \begin{equation} \phi(x_i,Q) = \int^Q
\Pi^{n-1}_{i=1} d^2 k_{\perp i} ~ \psi_{\rm val}(x_i,k_\perp).
\end{equation}
The distribution amplitudes are gauge-invariant vacuum to hadron
matrix elements and they obey evolution equation as dictated by the
OPE. Leading-twist PQCD predictions for hard exclusive
amplitudes~\cite{Lepage:1980fj}  are written  in a factorized form
as the product of hadron distribution amplitudes $\phi_I(x_i,Q)$ for
each hadron $I$ convoluted with  the hard scattering amplitude $T_H$
obtained by replacing each hadron with collinear on-shell quarks
with light-front momentum fractions $x_i = k^+_i/P^+.$  The
logarithmic evolution equations for the distribution amplitudes
require that the valence light-front wavefunctions fall-off
asymptotically as  the nominal power $[{1\over k^2_\perp}]^{n-1}, $
where $n$ is the number of elementary fields in the minimal Fock
state.

The light-front Fock representation thus provides an exact
formulation of current matrix elements of local and bi-local
operators.  In contrast, in equal-time Hamiltonian theory, one
must evaluate connected time-ordered diagrams where the gauge
particle or graviton couples to particles associated with vacuum
fluctuations.  Thus even if one knows the equal-time wavefunction
for the initial and final hadron, one cannot determine the current
matrix elements.  In the case of the covariant Bethe-Salpeter
formalism, the evaluation of the matrix element of the current
requires the calculation of an infinite number of irreducible
diagram contributions.

\section{General Structure of Light-Front Wavefunctions}

Even without explicit solutions, much is known about the explicit
form and structure of LFWFs.  They can  be matched to
nonrelativistic Schrodinger wavefunctions at soft scales. At high
momenta, the LFWFs at large $k_\perp$ and $x_i \to 1$ are
constrained by arguments based on conformal symmetry, the operator
product expansion, or perturbative QCD.  The pattern of higher Fock
states with extra gluons is given by ladder
relations~\cite{Antonuccio:1997tw}.

The structure of Fock states with nonzero orbital angular momentum is also constrained by
the Karmanov-Smirnov operator~\cite{ks92}. One can define the light-front Fock expansion
using a covariant generalization of light-front time: $\tau=x \cd \omega$. The
four-vector $\omega$, with $\omega^2 = 0$, determines the orientation of the light-front
plane; the freedom to choose $\omega$ provides an explicitly covariant formulation of
light-front quantization~\cite{cdkm}: all observables such as matrix elements of local
current operators, form factors, and cross sections are light-front invariants -- they
must be independent of $\omega_\mu.$ In recent work, Dae Sung Hwang, John Hiller, Volodya
Karmonov and I~\cite{Brodsky:2003pw} have studied the analytic structure of LFWFs using
the explicitly Lorentz-invariant formulation of the front form.  Eigensolutions of the
Bethe-Salpeter equation have specific angular momentum as specified by the Pauli-Lubanski
vector.  The corresponding LFWF for an $n$-particle Fock state evaluated at equal
light-front time $\tau = \omega\cdot x$ can be obtained by integrating the Bethe-Salpeter
solutions over the corresponding relative light-front energies.  The resulting LFWFs
$\psi^I_n(x_i, k_{\perp i})$ are functions of the light-front momentum fractions $x_i=
{k_i\cdot \omega / p \cdot \omega}$ and the invariant mass  of the constituents
$\mathcal{M}_n,$ each multiplying spin-vector and polarization tensor invariants which
can involve $\omega^\mu.$  They are eigenstates of the Karmanov--Smirnov kinematic
angular momentum operator~\cite{ks92,cdkm}.
\begin{equation}\label{ac1}
\vec{J} = -i[\vec{k}\times
\partial/\partial\vec{k}\,]-i[\vec{n}\times
\partial/\partial\vec{n}] +\frac{1}{2}\vec{\sigma},
\end{equation}
where $\vec n$ is the spatial component of $\omega$ in the
constituent rest frame ($\vec{\mathcal{P}}=\vec 0$).  Although this
form is written specifically in the constituent rest frame, it can
be generalized to an arbitrary frame by a Lorentz boost.

Normally the generators of angular rotations in the LF formalism
contain interactions, as in the Pauli--Lubanski formulation;
however, the LF angular momentum operator can also be represented in
the kinematical form (\ref{ac1}) without interactions. The key term
is the generator of rotations of the LF plane
$-i[\vec{n}\times\partial/\partial\vec{n}]$ which replaces the
interaction term; it appears only in the explicitly covariant
formulation, where the dependence on $\vec{n}$ is present. Thus
LFWFs satisfy all Lorentz symmetries of the front form, including
boost invariance, and they are proper eigenstates of angular
momentum.

In principle, one can solve for the LFWFs directly from the
fundamental theory using methods such as discretized light-front
quantization (DLCQ)~\cite{Pauli:1985ps}, the transverse
lattice~\cite{Bardeen:1979xx,Dalley:2004rq,Burkardt:2001jg}, lattice
gauge theory moments~\cite{DelDebbio:1999mq}, Dyson-Schwinger
techniques~\cite{Maris:2003vk}, and Bethe--Salpeter
techniques~\cite{Brodsky:2003pw}. DLCQ has been remarkably
successful in determining the entire spectrum and corresponding
LFWFs in one space-one time field theories~\cite{Gross:1997mx},
including QCD(1+1)~\cite{Hornbostel:1988fb} and
SQCD(1+1)~\cite{Harada:2004ck}. There are also DLCQ solutions for
low sectors of Yukawa theory in physical space-time
dimensions~\cite{Brodsky:2002tp}. The DLCQ boundary conditions allow
a truncation of the Fock space to finite dimensions while retaining
the kinematic boost and Lorentz invariance of light-front
quantization.

One can also project known solutions of the Bethe--Salpeter equation
to equal light-front time, thus producing hadronic light-front Fock
wave functions~\cite{Brodsky:2003pw}. Bakker and van Iersel have
developed new methods to find solutions to bound-state light-front
equations in ladder approximation~\cite{vanIersel:2004gf}. Pauli has
shown how one can construct an effective light-front Hamiltonian
which acts within the valence Fock state sector
alone~\cite{Pauli:2003tb}. Another possible  method is to construct
the $q\bar q$ Green's function using light-front Hamiltonian theory,
DLCQ boundary conditions and Lippmann-Schwinger resummation.  The
zeros of the resulting resolvent projected on states of specific
angular momentum $J_z$ can then generate the meson spectrum and
their light-front Fock wavefunctions.  As emphasized by Weinstein
and Vary, new effective operator
methods~\cite{Weinstein:2004nr,Zhan:2004ct} which have been
developed for Hamiltonian theories in  condensed matter and nuclear
physics, could also be applied advantageously to light-front
Hamiltonian.  Reviews of nonperturbative light-front methods may be
found in
references~\cite{Brodsky:1997de,cdkm,Dalley:ug,Brodsky:2003gk}.

Other important nonperturbative QCD methods are Dyson-Schwinger
techniques~\cite{Maris:2003vk} and the transverse
lattice~\cite{Dalley:2004rq}.  The transverse lattice method
combines DLCQ for one-space and the light-front time dimensions with
lattice theory in transverse space. It has recently provided the
first computation of the generalized parton distributions of the
pion~\cite{Dalley:2004rq}.

Currently the most important computational tool for making predictions in strong-coupling
QCD(3+1) is lattice gauge theory~\cite{Wilson:1976zj} which has made enormous progress in
recent years, particularly in computing mass spectra and decay constants. Lattice gauge
theory can only provide limited dynamical information because of the difficulty of
continuing predictions from Euclidean to Minkowski space. At present, results are limited
to large quark and pion masses such that the $\rho$ meson is
stable~\cite{DeGrand:2003xu}. In contrast to lattice gauge theory path integral methods,
Light-front Hamiltonian methods are frame-independent, formulated in Minkowski space,
only  two physical polarization gluonic degrees of freedom appear as quanta,  and there
is no complications from fermions. The known DLCQ solutions for 1+1 quantum field
theories could provide a powerful test of lattice methods.

The Hamiltonian approach is in fact the method of choice in
virtually every area of physics and quantum chemistry.  It has the
desirable feature that the output of such a calculation is
immediately useful: the spectrum of states and wavefunctions.
Furthermore, it allows the use of intuition developed in the study
of simple quantum systems, and also the application of, {\em e.g.},
powerful variational techniques.  The one area of physics where it
is {\em not} widely employed is relativistic quantum field theory.
The basic reason for this is that in a relativistic field theory
quantized at equal time (``the Instant Form") one has particle
creation/annihilation in the vacuum.  Thus the true ground state is
in general extremely complicated, involving a superposition of
states with arbitrary numbers of bare quanta, and one must
understand the complicated structure of this state before
excitations can be considered.  Furthermore, one must have a
nonperturbative way of separating out disconnected contributions to
physical quantities, which are physically irrelevant.  Finally, the
truncations that are required inevitably violate Lorentz covariance
and, for gauge theories, gauge invariance. These difficulties (along
with the development of covariant Lagrangian techniques) eventually
led to the almost complete abandonment of fixed-time Hamiltonian
methods in relativistic field theories.

Light-front quantization provides an alternative to the usual
formulation of field theories in which these problems appear to be
tractable.  This raises the prospect of developing a practical
Hamiltonian approach to solving field theories nonperturbatively
based on diagonalizing LC Hamiltonians.

\section{Consequences of Near-Conformal Field Theory}

One of the most exciting recent developments is the AdS/CFT
correspondence~\cite{Maldacena:1997re,Polchinski:2001tt,%
Brower:2002er,Andreev:2002aw} between superstring theory in 10
dimensions and supersymmetric Yang Mills theory in 3+1 dimensions.
As I will discuss below, one can use this connection to establish
the form of QCD wavefunctions at large transverse momentum
$k^2_\perp \to \infty$ and at $x \to 1$~\cite{Brodsky:2003px}.  The
AdS/CFT correspondence has important implications for hadron
phenomenology in the conformal limit, including an all-orders
demonstration of counting
rules~\cite{Brodsky:1973kr,Matveev:1973ra,Brodsky:1974vy} for hard
exclusive processes~\cite{Polchinski:2001tt}, as well as determining
essential aspects of hadronic light-front
wavefunctions~\cite{Brodsky:2003px}.

\subsection{The Conformal Correspondence Principle}

The classical Lagrangian of QCD for massless quarks is conformally
symmetric.  Since it has no intrinsic mass scale, the classical
theory is invariant under the $SO(4,2)$ translations, boosts, and
rotations of the Poincare  group, plus the dilatations and other
transformations of the conformal group. Scale invariance and
therefore conformal symmetry is destroyed in the quantum theory by
the renormalization procedure which introduces a renormalization
scale as well as by quark masses. Conformal symmetry is thus
broken in physical QCD; nevertheless, we can still recover the
underlying features of the conformally invariant theory by
evaluating any expression in QCD in the analytic limit of zero
quark mass and zero $\beta$ function~\cite{Parisi:zy}:
\begin{equation}
\lim_{m_q \to 0, \beta \to 0} \mathcal{O}_{QCD} = \mathcal{ O}_{\rm
conformal\ QCD} \ .
\end{equation} This conformal correspondence limit is analogous
to Bohr's correspondence principle where one recovers predictions
of classical theory from quantum theory in the limit of zero
Planck constant.  The contributions to an expression in QCD from
its nonzero $\beta$-function can be systematically
identified~\cite{Brodsky:2000cr,Rathsman:2001xe,Grunberg:2001bz}
order-by-order in perturbation theory using the Banks-Zaks
procedure~\cite{Banks:1981nn}.

The ``conformal correspondence principle" provides a new tool, the
conformal template~\cite{Brodsky:2004qb,Brodsky:2003dn} , which is
very useful for theory analyses, such as the expansion polynomials
for distribution
amplitudes~\cite{Brodsky:1980ny,Brodsky:1984xk,Brodsky:1985ve,Braun:2003rp},
the non-perturbative wavefunctions which control exclusive processes
at leading twist~\cite{Lepage:1979zb,Brodsky:2000dr}.

\subsection{Commensurate Scale Relations}

The near-conformal behavior of QCD is the basis for commensurate scale
relations~\cite{Brodsky:1994eh} which relate observables to each other without
renormalization scale or scheme ambiguities~\cite{Brodsky:2000cr,Rathsman:2001xe}. One
can derive the commensurate scale relation between the effective charges of any two
observables by first computing their relation in conformal gauge theory; the effects of
the nonzero QCD $\beta-$ function are then taken into account using the BLM
method~\cite{Brodsky:1982gc} to set the scales of the respective couplings. An important
example is the generalized Crewther relation~\cite{Brodsky:1995tb}:
\begin{equation}
\left[1 + \frac{\alpha_R(s^*)}{\pi} \right] \left[1 -
\frac{\alpha_{g_1}(Q^2)}{\pi}\right] = 1
\end{equation}
where the underlying form at zero $\beta$ function is dictated by conformal
symmetry~\cite{Crewther:1972kn}. Here $\alpha_R(s)/\pi$ and $-\alpha_{g_1}(Q^2)/\pi$
represent the entire radiative corrections to $R_{e^+ e^-}(s)$ and the Bjorken sum rule
for the $g_1(x,Q^2)$ structure function measured in spin-dependent deep inelastic
scattering, respectively. The relation between $s^*$ and $Q^2$ can be computed order by
order in perturbation theory using the BLM method~\cite{Brodsky:1982gc}.   The ratio of
physical scales guarantees that the effect of new quark thresholds is commensurate.
Commensurate scale relations are renormalization-scheme independent and satisfy the group
properties of the renormalization group.  Each observable can be computed in any
convenient renormalization scheme such as dimensional regularization. The $\bar{MS}$
coupling can then be eliminated; it becomes only an intermediary~\cite{Brodsky:1994eh}.
In such a procedure there are no further renormalization scale ($\mu$) or scheme
ambiguities.

The  effective charge~\cite{Brodsky:1997dh} defined from the ratio of elastic pion and
photon-to-pion transition form factors $\alpha^{\rm exclusive}_s(Q^2) = {F_\pi(Q^2)/ 4\pi
Q^2 F^2_{\gamma \pi^0}(Q^2)}$ can also be connected to other effective charges and
observables by commensurate scale relations. Its magnitude, $\alpha^{\rm
exclusive}_s(Q^2) \sim 0.8$ at small $Q^2,$  is sufficiently large as to explain the
observed magnitude of exclusive amplitudes such as the pion form factor using the
asymptotic distribution amplitude. An analytic effective charge such as the pinch
scheme~\cite{Cornwall:1981zr} provides a method to unify the electroweak and strong
couplings and forces.

\subsection{ Fixed Point Behavior}

Although the QCD coupling decreases logarithmically at high
virtuality  due to asymptotic freedom, theoretical~\cite{vonSmekal:1997is,Zwanziger:2003cf,%
Howe:2002rb,Howe:2003mp,Furui:2003mz,Furui:2004bq,
Badalian:2004ig,Ackerstaff:1998yj} and
phenomenological~\cite{Mattingly:ej,Brodsky:2002nb,Baldicchi:2002qm}
evidence is now accumulating that the QCD coupling becomes constant
at small virtuality; {\em i.e.}, $\alpha_s(Q^2)$ develops an
infrared fixed point in contradiction to the usual assumption of
singular growth in the infrared.  If QCD running couplings are
bounded, the integration over the running coupling is finite and
renormalon resummations are  not required.  If the QCD coupling
becomes scale-invariant in the infrared, then elements of conformal
theory~\cite{Braun:2003rp} become relevant even at relatively small
momentum transfers.

Menke, Merino, and Rathsman and I have presented a definition of a
physical coupling for QCD which has a direct relation to high
precision measurements of the hadronic decay channels of the $\tau^-
\to \nu_\tau {\rm H}^-$~\cite{Brodsky:2002nb} .  Let $R_{\tau}$ be
the ratio of the hadronic decay rate to the leptonic one.  Then
$R_{\tau}\equiv R_{\tau}^0\left[1+\frac{\alpha_\tau}{\pi}\right]$,
where $R_{\tau}^0$ is the zeroth order QCD prediction, defines the
effective charge $\alpha_\tau$.  The data for $\tau$ decays is
well-understood channel by channel, thus allowing the calculation of
the hadronic decay rate and the effective charge as a function of
the $\tau$ mass below the physical mass.  The vector and
axial-vector decay modes can be studied separately. Using an
analysis of the $\tau$ data from the OPAL
collaboration~\cite{Ackerstaff:1998yj}, we have found that the
experimental value of the coupling $\alpha_{\tau}(s)=0.621 \pm
0.008$ at $s = m^2_\tau$ corresponds to a value of
$\alpha_{\MSbar}(M^2_Z) = (0.117$-$0.122) \pm 0.002$, where the
range corresponds to three different perturbative methods used in
analyzing the data.  This result is in good agreement with the world
average $\alpha_{\MSbar}(M^2_Z) = 0.117 \pm 0.002$.  However, one
also finds that the effective charge only reaches $\alpha_{\tau}(s)
\sim 0.9 \pm 0.1$ at $s=1\,{\rm GeV}^2$, and it even stays within
the same range down to $s\sim0.5\,{\rm GeV}^2$. The effective
coupling is close to constant at low scales, suggesting that
physical QCD couplings become constant or ``frozen" at low scales.

\begin{figure}[htb]
\centering
\includegraphics[width=4.3in]   
{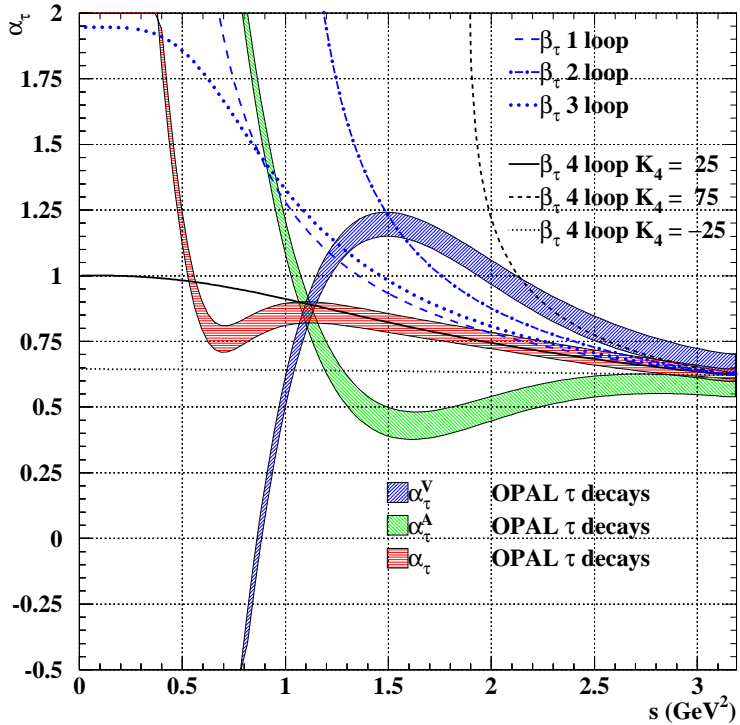} \caption[*]{The effective charge $\alpha_{\tau}$ for non-strange hadronic
decays of a hypothetical $\mathit{\tau}$ lepton with $\mathit{m_{\tau'}^2 = s}$ compared
to solutions of the fixed order evolution equation for $\alpha_{\tau}$ at two-, three-,
and four-loop order.  The error bands include statistical and systematic errors.
\label{fig:fopt_comp}}
\end{figure}

Figure~\ref{fig:fopt_comp} shows a comparison of the experimentally determined effective
charge $\alpha_{\tau}(s)$ with solutions to the evolution equation for $\alpha_{\tau}$ at
two-, \hbox{three-,} and four-loop order normalized at $m_\tau$.  At three loops the
behavior of the perturbative solution drastically changes, and instead of diverging, it
freezes to a value $\alpha_{\tau}\simeq 2$ in the infrared. The infrared behavior is not
perturbatively stable since the evolution of the coupling is governed by the highest
order term.  This is illustrated by the widely different results obtained for three
different values of the unknown four loop term $\beta_{\tau,3}$ which are also shown. The
values of $\beta_{\tau,3}$ used are obtained from the estimate of the four loop term in
the perturbative series of $R_\tau$, $K_4^{\overline{\rm MS}} = 25\pm
50$~\cite{LeDiberder:1992fr}. It is interesting to note that the central four-loop
solution is in good agreement with the data all the way down to $s\simeq1\,{\rm GeV}^2$.

The results for $\alpha_{\tau}$ resemble the behavior of the one-loop ``time-like"
effective coupling~\cite{Beneke:1994qe,Ball:1995ni,Dokshitzer:1995qm}
\begin{equation}\label{eq:alphaeff}
\alpha_{\rm eff}(s)=\frac{4\pi}{\beta_0} \left\{\frac{1}{2} -
\frac{1}{\pi}\arctan\left[\frac{1}{\pi}\ln\frac{s}{\Lambda^2}\right]\right\}
\end{equation}
which is finite in the infrared and freezes to the value $\alpha_{\rm
eff}(s)={4\pi}/{\beta_0}$ as $s\to 0$.  It is instructive to expand the ``time-like"
effective coupling for large $s$,
\begin{eqnarray}
\alpha_{\rm eff}(s) &=&\frac{4\pi}{\beta_0\ln\left(s/\Lambda^2\right)} \left\{1
-\frac{1}{3}\frac{\pi^2}{\ln^2\left(s/\Lambda^2\right)}
+\frac{1}{5}\frac{\pi^4}{\ln^4\left(s/\Lambda^2\right)} +\ldots \right\} \nonumber\\
&=&\alpha_{\rm s}(s)\left\{1 -\frac{\pi^2\beta_0^2}{3}\left(\frac{\alpha_{\rm
s}(s)}{4\pi}\right)^2 +\frac{\pi^4\beta_0^4}{5}\left(\frac{\alpha_{\rm
s}(s)}{4\pi}\right)^4 +\ldots \right\}.
\end{eqnarray}
This shows that the ``time-like" effective coupling is a resummation of
$(\pi^2\beta_0^2\alpha_{\rm s}^2)^n$-corrections to the usual running couplings.  The
finite coupling $\alpha_{\rm eff}$ given in Eq.~(\ref{eq:alphaeff}) obeys standard PQCD
evolution at LO.  Thus one can have a solution for the perturbative running of the QCD
coupling which obeys asymptotic freedom but does not have a Landau singularity.

The near constancy of the effective QCD coupling at small scales
helps explain the empirical success of dimensional counting rules
for the power law fall-off of form factors and fixed angle scaling.
As shown in the references~\cite{Brodsky:1997dh,Melic:2001wb}, one
can calculate the hard scattering amplitude $T_H$ for such
processes~\cite{Lepage:1980fj} without scale ambiguity in terms of
the effective charge $\alpha_\tau$ or $\alpha_R$ using commensurate
scale relations.  The effective coupling is evaluated in the regime
where the coupling is approximately constant, in contrast to the
rapidly varying behavior from powers of $\alpha_{\rm s}$ predicted
by perturbation theory (the universal two-loop coupling).  For
example, the nucleon form factors are proportional at leading order
to two powers of $\alpha_{\rm s}$ evaluated at low scales in
addition to two powers of $1/q^2$; The pion photoproduction
amplitude at fixed angles is proportional at leading order to three
powers of the QCD coupling.  The essential variation from
leading-twist counting-rule behavior then only arises from the
anomalous dimensions of the hadron distribution amplitudes.

\subsection{ The Abelian Correspondence Principle}

Another important guide to QCD predictions is consistency in
a limit where the theory becomes Abelian. One can consider QCD
predictions as functions of analytic variables of the number of
colors $N_C$ and flavors $N_F$. At $N_C \to \infty$ at fixed $N_C
\alpha_s,$ calculations in QCD greatly simplify since only planar
diagrams enter. However, the $N_C \to 0$ limit is also very
interesting. Remarkably, one can show at all orders of
perturbation theory~\cite{Brodsky:1997jk} that PQCD predictions
reduce to those of an Abelian theory similar to QED at $N_C \to 0$
with $C_F \alpha_s$ and $N_F\over T_F C_F$ held fixed, where
$C_F={N^2_C-1\over 2 N_C}$ and $T_F=1/2.$ The resulting theory
corresponds to the group ${1/U(1)}$ which means that
light-by-light diagrams acquire a particular topological factor.
The $N_C \to 0$ limit provides an important check on QCD analyses;
QCD formulae and phenomena must match their Abelian analog. The
renormalization scale is effectively fixed by this requirement.
Commensurate scale relations obey the Abelian Correspondence
principle, giving the correct Abelian relations between
observables in the limit $N_C \to 0.$

\section{Perturbative QCD and Exclusive Processes}

Exclusive processes constitute provide an important window on QCD
processes and the  structure of hadrons. There has been considerable
progress analyzing exclusive and diffractive reactions at large
momentum transfer from first principles in QCD. Rigorous statements
can be made on the basis of asymptotic freedom and factorization
theorems which separate the underlying hard quark and gluon
subprocess amplitude from the nonperturbative physics of the
hadronic wavefunctions. The leading-power contribution to exclusive
hadronic amplitudes such as quarkonium decay, heavy hadron decay,
and scattering amplitudes where hadrons are scattered with large
momentum transfer can often be factorized as a convolution of
distribution amplitudes $\phi_H(x_i,\Lambda)$ and hard-scattering
quark/gluon scattering amplitudes $T_H$ integrated over the
light-front momentum fractions of the valence
quarks~\cite{Lepage:1980fj}:
\begin{eqnarray}
\M_{\rm Hadron} &=&\int
 \prod \phi_H^{(\Lambda)} (x_i,\lambda_i)\, T_H^{(\Lambda)} dx_i\ .
\label{eq:e}
\end{eqnarray}
Here $T_H^{(\Lambda)}$ is the underlying quark-gluon subprocess
scattering amplitude in which each incident and final hadron is
replaced by valence quarks with collinear momenta $k^+_i =x_i
p^+_H$, $\vec k_{\perp i} = x_i \vec p_{\perp H }.$ The invariant
mass of all intermediate states in $T_H$ is evaluated above the
separation scale $\M^2_n > \Lambda^2$. The essential part of the
hadronic wavefunction is the distribution
amplitude~\cite{Lepage:1980fj}, defined as the integral over
transverse momenta of the valence (lowest particle number) Fock
wavefunction; \eg\ for the pion
\begin{equation}
\phi_\pi (x_i,Q) \equiv \int d^2k_\perp\, \psi^{(Q)}_{q\bar q/\pi} (x_i,
\vec k_{\perp
i},\lambda) \label{eq:f}
\end{equation}
where the separation scale $\Lambda$ can be taken to be order of
the characteristic momentum transfer $Q$ in the process. It should
be emphasized that the hard scattering amplitude $T_H$ is
evaluated in the QCD perturbative domain where the propagator
virtualities are above the separation scale.

The leading power fall-off of the hard scattering amplitude as given
by dimensional counting rules follows from the nominal scaling of
the hard-scattering amplitude: $T_H \sim 1/Q^{n-4}$, where $n$ is
the total number of fields (quarks, leptons, or gauge fields)
participating in the hard
scattering~\cite{Brodsky:1974vy,Matveev:1973ra}. Thus the reaction
is dominated by subprocesses and Fock states involving the minimum
number of interacting fields.  In the case of $2 \to 2$ scattering
processes, this implies
\begin{equation}
{d\sigma\over dt}(A B \to C D) ={F_{A B \to C D}(t/s)/ s^{n-2}}
\end{equation}
where $n = N_A + N_B + N_C +N_D$ and $n_H$ is the minimum number of
constituents of $H$.

In the case of form factors, the dominant helicity conserving
amplitude has the nominal power-law falloff  $F_H(t) \sim
(1/t)^{n_H-1},$ The complete predictions from PQCD modify the
nominal scaling by logarithms from the running coupling and the
evolution of the distribution amplitudes. In some cases, such as
large angle $pp \to p p $ scattering, there can be ``pinch"
contributions~\cite{Landshoff:ew} when the scattering can occur from
a sequence of independent near-on shell quark-quark scattering
amplitudes at the same CM angle.  After inclusion of Sudakov
suppression form factors, these contributions also have a scaling
behavior close to that predicted by constituent counting.

The constituent counting rules were originally derived in
1973~\cite{Brodsky:1974vy,Matveev:1973ra} before the development of
QCD in anticipation that the underlying theory of hadron physics
would be renormalizable and close to a conformal theory.  The
factorized structure of hard exclusive amplitudes in terms of a
convolution of valence hadron wavefunctions times a hard-scattering
quark scattering amplitude was also proposed~\cite{Brodsky:1974vy}.
Upon the discovery of the asymptotic freedom in QCD, there was a
systematical development of the theory of hard exclusive reactions,
including factorization theorems, counting rules, and evolution
equations for the hadronic distribution
amplitudes~\cite{Brodsky:1979qm,Lepage:1979za,Lepage:1979zb,Efremov:1980rn}.

The distribution amplitudes which control leading-twist exclusive
amplitudes at high momentum transfer can be related to the
gauge-invariant Bethe-Salpeter wavefunction at equal light-front
time $\tau = x^+$.  The logarithmic evolution of the hadron
distribution amplitudes $\phi_H (x_i,Q)$ with respect to the
resolution scale $Q$ can be derived from the
perturbatively-computable tail of the valence light-front
wavefunction in the high transverse momentum regime. The DGLAP
evolution of quark and gluon distributions can also be derived in an
analogous way by computing the variation of the Fock expansion with
respect to the separation scale. Other key features of the
perturbative QCD analyses are: (a) evolution equations for
distribution amplitudes which incorporate the operator product
expansion, renormalization group invariance, and conformal
symmetry~\cite{Lepage:1980fj,Brodsky:1980ny,Muller:1994cn,%
Ball:1998ff,Braun:1999te}; (b) hadron helicity conservation which
follows from the underlying chiral structure of
QCD~\cite{Brodsky:1981kj}; (c) color transparency, which eliminates
corrections to hard exclusive amplitudes from initial and final
state interactions at leading power and reflects the underlying
gauge theoretic basis for the strong
interactions~\cite{Brodsky:1988xz} and (d) hidden color degrees of
freedom in nuclear wavefunctions, which reflect the color structure
of hadron and nuclear wavefunctions~\cite{Brodsky:1983vf}. There
have also been recent advances eliminating renormalization scale
ambiguities in hard-scattering amplitudes via commensurate scale
relations~\cite{Brodsky:1994eh} which connect the couplings entering
exclusive amplitudes to the $\alpha_V$ coupling which controls the
QCD heavy quark potential.

Exclusive processes such as $\bar p p \to \bar p p,$ $\bar p p \to
K^+ K^-$ and $\bar p p \to \gamma \gamma$ provide a unique window
for viewing QCD processes and hadron dynamics at the amplitude
level~\cite{Brodsky:1981rp,Brodsky:2000dr}.  New tests of theory
and comprehensive measurements of hard exclusive amplitudes can
also be carried out for electroproduction at Jefferson Laboratory
and in two-photon collisions at CLEO, Belle, and
BaBar~\cite{Brodsky:2001hv}.    Hadronic exclusive processes are
closely related to exclusive hadronic $B$ decays, processes which
are essential for determining the CKM phases and the physics of
$CP$ violation. The universal light-front wavefunctions which
control hard exclusive processes such as form factors, deeply
virtual Compton scattering, high momentum transfer
photoproduction, and two-photon processes, are also required for
computing exclusive heavy hadron
decays~\cite{Beneke:2000ry,Keum:2000wi,Szczepaniak:1990dt,Brodsky:2001jw},
such as $B \to K \pi$, $B \to \ell \nu \pi$, and $B \to K  p \bar
p$~\cite{Chua:2002wn}. The same physics issues, including color
transparency, hadron helicity rules, and the question of dominance
of leading-twist perturbative QCD mechanisms enter in both realms
of physics.

\begin{figure}[htb]
\centering
\includegraphics[width=4.3in]{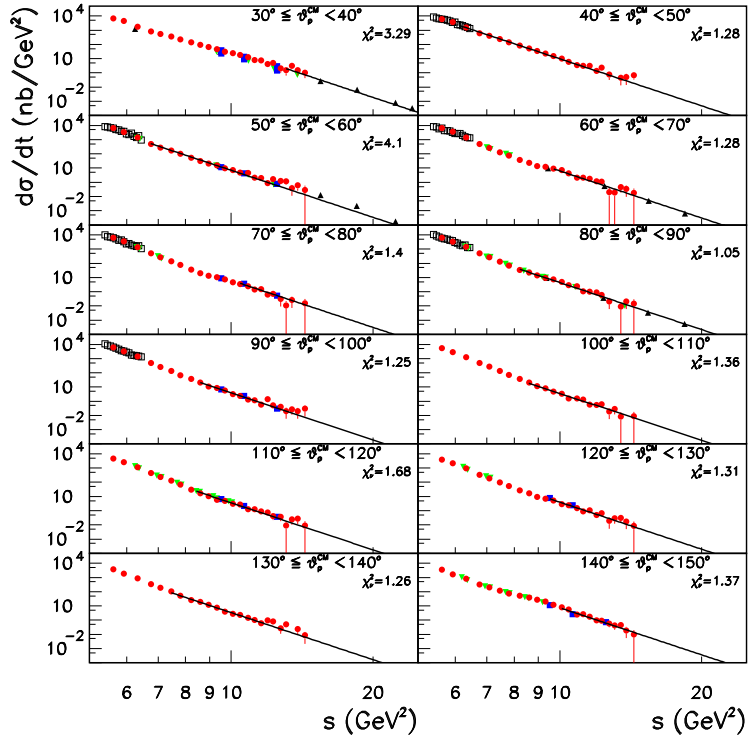}
\caption{Fits of the cross sections $d\sigma/dt$ to $s^{-11}$ for
$P_T \ge P_T^{th}$ and proton angles between $30^{\circ}$ and
$150^{\circ}$ (solid lines). Data are from CLAS (full/red circles),
Mainz(open/black squares), SLAC (full-down/green triangles), JLab
Hall~A (full/blue squares) and Hall~C (full-up/black triangles).
Also shown in each panel is the $\chi^2_\nu$ value of the fit. From
Ref.~\cite{Rossi:2004qm}.} \label{fig:fitfin}
\end{figure}

The data for virtually all measured hard scattering processes appear
to be consistent with the conformal predictions of QCD.
For example, one also sees the onset of the predicted
perturbative QCD scaling behavior for exclusive nuclear amplitudes
such as deuteron photodisintegration (Here $n = 1+ 6 + 3 + 3 = 13 .$)
$s^{11}{ d\sigma\over dt}(\gamma d \to p n) \sim $ constant at
fixed CM angle. The measured deuteron form factor and the deuteron
photodisintegration cross section appear to follow the
leading-twist QCD predictions at large momentum transfers in the
few GeV region~\cite{Holt:1990ze,Bochna:1998ca,Rossi:2004qm}. A
comparison of the data with the QCD predictions is shown in
Fig.~\ref{fig:fitfin}.

Another application to exclusive nuclear processes is the approach to scaling of the
deuteron form factor $[{Q^2}]^ 5\sqrt{A(Q^2)} \to {\rm const}$ observed at SLAC and
Jefferson laboratory at high $Q^2.$  These scaling laws reflects the underlying scaling
of the nucleon-nucleon interaction and the nuclear force at short distances. The
phenomenological successes provide further evidence for the dominance of leading-twist
quark-gluon subprocesses and the near conformal behavior of the QCD coupling. As
discussed above, the evidence that the running coupling has constant fixed-point
behavior, which together with BLM scale fixing, could help explain the near conformal
scaling behavior of the fixed-CM angle cross sections.  The angular distribution of hard
exclusive processes is generally consistent with quark interchange, as predicted from
large $N_C$ considerations.

\section{The Evolution of the Deuteron Distribution Amplitude and Hidden Color}

In this section I will review an analysis by Chueng Ji,  Peter
Lepage, and myself which shows how the asymptotic behavior of the
deuteron form factor at large momentum transfer and the evolution of
the deuteron six-quark distribution amplitude at short distances can
be  computed systematically as an expansion in
$\alpha_s(Q^2)$~\cite{Brodsky:1983vf}.  The results agree with the
operator product expansion as well as the conformal scaling  implied
by the AdS/CFT correspondence.  As we shall see, the QCD predictions
appear to be in remarkable agreement with experiment for $Q^2 \gsim$
1~GeV$^2$ particularly when expressed in terms of the deuteron
reduced form factor. This provides a good check on the six-quark
description of the deuteron at short distances as well as the scale
invariance of the elastic quark-quark scattering amplitude.  I will
also discuss how the dominance of the hidden color amplitudes at
short distances also provides an explanation for the repulsive
behavior of the nucleon-nucleon potential at small inter-nucleon
separation.

Hadronic form factors in QCD at large momentum transfer $Q^2 = \vec
q\,^2-q^2_0$ can be written in a factorized form where all
nonperturbative effects are incorporated into process-independent
distribution amplitudes $\phi_H(x_i,Q)$, computed from the equal
$\tau = t+z$, six-quark valence wave function at small relative
quark transverse separation $b^i_\perp \sim O(1/Q)$. The $x_i =
(k^0+k^3)_i/(p^0+ p^3)$ are the light-front longitudinal momentum
fractions with $\sum^n_{i=1}x_i=1$.  In the case of the deuteron,
only the six-quark Fock state needs to be considered for the purpose
of computing a hard scattering amplitude since in a physical gauge
any additional quark or gluon forced to absorb large momentum
transfer yields a power law suppressed contribution to the form
factor.  The deuteron form factor can then be written as a
convolution
\begin{equation}
F_d(Q^2) = \int^1_0 [dx][dy]\phi^\dag_d(y,Q)\, T_H^{6_q+\gamma
^*\rightarrow 6_q}(x,y,Q)\phi_d(x,Q) \ , \label{eq:1}
\end{equation}
where the hard scattering amplitude
\begin{equation}
T_H^{6_q+\gamma^*\rightarrow 6_q} =
\left[\frac{\alpha_s(Q^2)}{Q^2}\right]^5 t(x,y)  \left[ 1+
O(\alpha_s(Q^2))\right] \label{eq:2}
\end{equation}
gives the probability amplitude for scattering six quarks
collinear with the initial to the final deuteron momentum and
\begin{equation}
\phi_d (x_i,Q) \propto \int^{k_{\perp\, i}< Q} \left[ d\,^2
k_\perp\right]\, \psi_{qqq\,qqq}(x_i,\vec k_{\perp\, i})
\label{eq:3}
\end{equation}
gives the probability amplitude for finding the quarks with
longitudinal momentum fractions $x_i$ in the deuteron wavefunction
collinear up to the scale $Q$.  Because the coupling of the gauge
gluon is helicity-conserving and the fact that $\phi_d(x_i,Q)$ is
the $L_z=0$ projection of the deuteron wavefunction, hadron helicity
is conserved:  The dominant form factor corresponds to
$\sqrt A(Q^2)$; {\em i.e.}, $h = h^\prime = 0$.

The distribution amplitude $\phi_d(x_i,Q)$ is the basic deuteron
wave function which controls high momentum transfer exclusive
reactions in QCD.  The logarithmic $Q^2$ dependence Of $\phi_d$ is
determined by an evolution equation computed from perturbative
quark-quark scattering kernels at large momentum transfer, or
equivalently, by the operator product expansion at short distances
and the renormalization
group~\cite{Lepage:1980fj,Duncan:1979hi,Brodsky:1981kj}.

The QCD prediction for the leading helicity-zero deuteron form
factor then has the form~\cite{Brodsky:1976mn,Brodsky:1976rz}
\begin{equation}
F_d(q^2) = \left[\frac{\alpha_s(Q^2)}{Q^2}\right]^5 \sum_{m,n}
d_{mn} \left(\ell n\,
\frac{Q^2}{\Lambda^2}\right)^{-\gamma_n^d-\gamma_m^d}\, \left[1+ O
\left(\alpha_s(Q^2),\, \frac{m}{Q}\right)\right] \ , \label{eq:4}
\end{equation}
where the main dependence $[\alpha_s(Q^2)/Q^2]^5$ comes from the
hard-gluon exchange amplitude $T_H$.  The anomalous dimensions
$\gamma^d_n$ are calculated from the evolution equations for
$\psi_d(x_i,Q)$.

The evolution equation for six quark systems in which the
constituents have the light-front longitudinal momentum fractions
$x_i~(i=1,2,\ldots,6)$ can be obtained from a generalization of the
proton (three quark)
case~\cite{Lepage:1980fj,Duncan:1979hi,Brodsky:1981kj}. A nontrivial
extension is the calculation of the color factor, $C_d$, of six
quark systems.  Since in leading order only pair-wise interactions,
with transverse momentum Q, occur between quarks, the evolution
equation for the six-quark system becomes $\{[dy]=
\delta(1-\sum^6_{i=1} y_i)\, \prod^6_{i=1}dy_i$, $C_F=(n^2_c-1)/2n_c
= (4/3), \beta=11 - (2/3)n_f$, and $n_f$ is the effective number of
flavors)
\begin{equation}
\prod^6_{k=1} x_k
\left[\frac{\partial}{\partial\xi}+\frac{3C_F}{\beta}\right]
\widetilde\Phi (x_i,Q) = - \frac{C_d}{\beta} \int^1_0 [dy]\,
V(x_i,y_i)\widetilde\Phi(y_i,Q) \,,\label{eq:5}
\end{equation}
where the factor 3 in the square bracket comes from the
renormalization of the six quark field.  In Eq.~(\ref{eq:5}) we have
defined $\Phi(x_i,Q) = \prod^6_{k=1} x_k \widetilde\Phi(x_i,Q)$. The
evolution is in the variable
\begin{equation}
\xi(Q^2) =
\frac{\beta}{4\pi} \int^{Q^2}_{Q_0^2}
\frac{dk^2}{k^2}\,
\alpha_s(k^2) \sim \ell n
\left(
\frac{\ell n
\frac{Q^2}{\Lambda^2}}
{\ell n
\frac{Q_0^2}{\Lambda^2}} \right) \ .
\label{eq:6}
\end{equation}
By summing over interactions between quark pairs $\{i, j\}$ due to
exchange of a single gluon, $V(x_i, y_i) = V(y_i, x_i)$ is given by
\begin{equation}
V(x_i,y_i) = 2 \prod^6_{k=1} x_k \sum^6_{i\ne j}
\theta(y_i-x_i) \prod^6_{\ell \ne i,j} \delta(x_\ell-y_\ell)\,
\frac{y_j}{x_j} \left(\frac{\delta_{h_i\bar h_j}}{x_i+x_j} +
\frac{\Delta}{y_i-x_i}\right) \ ,
\label{eq:7}
\end{equation}
where $\delta_{h_i\bar h_j}= 1(0)$ when the constituents' $\{i,j\}$
helicities are antiparallel (parallel).  The infrared singularity at
$x_i = y_i$ is cancelled by the factor $\Delta\widetilde\Phi(y_i,Q)
= \widetilde\Phi(y_i,Q)-\widetilde\Phi(x_i,Q)$ since the deuteron is
a color singlet.

The six-quark bound states have five independent color singlet
components $(3 \times 3 \times 3 \times 3 \times 3 \times 3 \supset
\underline{\,1}+\underline{\,1}+\underline{\,1}+\underline{\,1}+\underline{\,1})$.
It can be shown in general that the color factor $C_d$ is given by
\begin{equation}
C_d = \frac{1}{5}\, S_{ijk\ell mn}{}^\alpha
\left(\frac{1}{2} \lambda_a\right)_{i^\prime}
{}^i\left(\frac{1}{2} \lambda_a\right)_{j^\prime}
{}^jS_\alpha^{i^\prime j^\prime k \ell mn} \ ,
\label{eq:8}
\end{equation}
where $\lambda_a(a=1,2,\ldots,8)$ are Gell-Mann matrices in
$SU(3)^c$ group and \break $s_{ijk\ell
mn}^\alpha(\alpha=1,2,\ldots,5)$ are the five independent color
singlet representations.  We shall focus on results for the leading
contribution to the distribution amplitude and form factor at large
$Q$.  Since the leading eigensolution to the evolution
Eq.~(\ref{eq:5}) turns out to be completely symmetric in its orbital
dependence, the dominant asymptotic deuteron wavefunction is fixed
by overall anti-symmetry  to have spin-isospin symmetry $\{3\}_{TS}$
which is dual to its color symmetry [222]$_c$. Thus the coefficient
for each $c$ (and $TS$) component has equal weights:
\begin{equation}
\phi_{6_q} \left([222]_c \otimes \{33\}_{TS}\right) =
\frac{1}{\sqrt 5} \sum^5_{\alpha=1} (-1)^\alpha
[222]_c {} ^\alpha\{33\}_{TS}^\alpha \ .
\label{eq:9}
\end{equation}
Since the evolution potential is diagonal in isospin and spin, $C_d$
is computed by the trace of the color representation.  The color
factor is $-2/3$ for the color antisymmetric pair $\{i,j\}$ and
$+1/3$ for the color symmetric pair $\{i,j\}$.  Since three color
antisymmetric pairs $\{i,j\}$ and two color symmetric pairs
 $\{i,j\}$ exist in this state, the color factor is
\begin{equation}
C_d = \frac{1}{5} \left( -\frac{2}{3} \times 3 + \frac{1}{3} \times 2\right) =
\frac{C_F}{5} \ .
\label{eq:10}
\end{equation}
To solve the evolution Eq.~(\ref{eq:5}), we factorize the $Q^2$
dependence of $\widetilde\Phi(x_i,Q)$ as
\begin{equation}
\widetilde\Phi (x_i,Q) = \widetilde\Phi(x_i)\, e^{-\gamma\xi} =
\widetilde\Phi(x_i)\, \left[\ell n
\frac{Q^2}{\Lambda^2}\right]^{-\gamma} \ , \label{eq:11}
\end{equation}
where the eigenvalues of $\gamma$ will provide the anomalous
dimensions $\gamma_n$.
The leading anomalous dimension $\gamma_0$ [corresponding
 to the eigenfunction $\widetilde\Phi(x_i)=1$] is
\begin{equation}
\gamma_0 = \frac{3C_F}{\beta} + \frac{C_d}{\beta} \sum^6_{i\ne j}
\delta_{h_i \bar h_j} \ ,
\label{eq:12}
\end{equation}
so that the asymptotically dominant result for the helicity zero
deuteron is given by $\gamma_0 = (6/5)(C_F/\beta)$.

Note that in order to have  logarithmic evolution of the deuteron
distribution amplitude, the six-quark valence light-front
wavefunction must fall nominally as \break
$\psi_{qqqqqq/d}(x_i,k_{\perp i} )\simeq  [ {1\over k_\perp^2}]^5.$
This is also the prediction of conformal invariance and the AdS/CFT
correspondence.  More generally, consistency with the operator product expansion for the
moments of the distribution amplitude requires the power law fall off
$\psi_{n}(x_i,k_{\perp i} )\simeq  [ {1\over k_\perp^2}]^{n-1}$ for all $n$-parton
LFWFs with $L_z=0.$

At high $Q^2$ the deuteron form factor is sensitive to wavefunction
configurations where all six quarks overlap within an impact
separation $b_{\perp i} < \mathcal{O} (1/Q).$
Since the deuteron
form factor contains the probability amplitudes for the proton and
neutron to scatter from $p/2$ to $p/2+q/2$, it is natural to define
the reduced deuteron form factor\cite{Brodsky:1976rz,Brodsky:1983vf}
\begin{equation} f_d(Q^2) \equiv {F_d(Q^2)\over
F_{1N} \left(Q^2\over 4\right)\, F_{1N}\,\left(Q^2\over
4\right)}.
\label{eq:13}
\end{equation}
The effect of nucleon compositeness is
removed from the reduced form factor.
 Since the
leading anomalous dimensions of the nucleon distribution amplitude
is $C_F/2\beta$, the QCD prediction for the asymptotic $Q^2$
behavior of $f_d(Q^2)$ is
\begin{equation}
f_d(Q^2) \sim \frac{\alpha_s(Q^2)}{Q^2} \, \left(\ell n
\frac{Q^2}{\Lambda^2}\right)^{(2/5)\, C_F/\beta} \ ,
\label{eq:14}
\end{equation}
where  $(2/5)(C_F/\beta) = - (8/145)$ for $n_f=2$.

\begin{figure}[htb]
\begin{center}
\includegraphics[width=4in,height=2.5in]{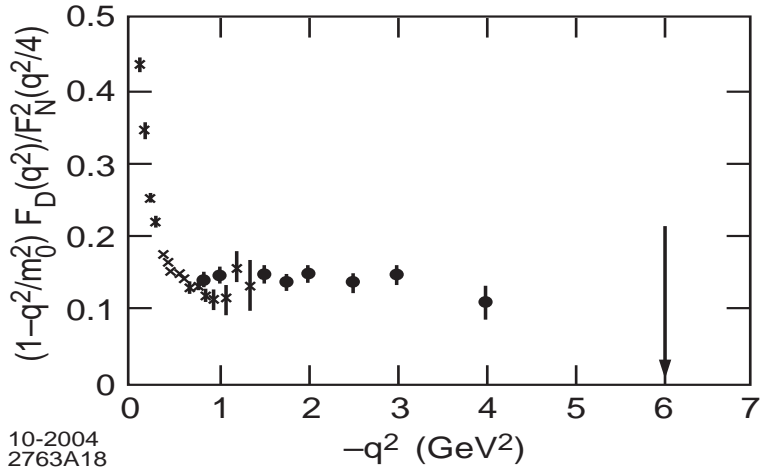}\end{center}
\caption[*]{Reduced Deuteron Form Factor showing the scaling
predicted by perturbative QCD and conformal scaling.  The data show
two regimes: a fast-falling behavior at small $Q^2$ characteristic
of normal nuclear binding, and a hard scattering regime with
monopole fall-off controlled by the scale $m^2_0 = 0.28~{\rm
GeV^2}.$ The latter contribution is attributable to non-nucleonic
hidden-color components of the deuteron's six-quark Fock state. From
Ref.~\cite{Brodsky:1976rz}. \label{reduced}}
\end{figure}

QCD thus predicts essentially the  same scaling law for the reduced
deuteron form factor as a meson form factor.  This scaling is
consistent with experiment for $Q^2 > 1~{\rm GeV}^2.$  In fact as
seen in Fig.~\ref{reduced}, the deuteron reduced form factor
contains two components: (1) a fast-falling component characteristic
of nuclear binding with probability $85\%$, and (2) a hard
contribution falling as a monopole with a scale of order $0.5~{\rm
GeV}$  with probability $15\%.$ The normalization of the deuteron
form factor observed at large $Q^2$~\cite{Arnold:1975dd}, as well as
the presence of two mass scales in the scaling behavior of the
reduced deuteron form factor~\cite{Brodsky:1976rz} thus suggests
sizable hidden-color Fock state contributions such as
$\ket{(uud)_{8_C} (ddu)_{8_C}}$ with probability  of order $15\%$ in
the deuteron wavefunction~\cite{Farrar:1991qi}.

In general, one would expect corrections from the leading twist QCD
predictions from higher-twist effects ({\em e.g.}, mass and
$k_\perp$ smearing) and higher-order contributions in
$\alpha_s(Q^2)$, as well as nonleading anomalous dimensions.
However, the agreement of the data with simple $Q^2f_d(Q^2)\sim$
const. behavior for $Q^2> 1/2$ GeV$^2$ implies that, unless there is
a fortuitous cancellations, all of the scale-breaking effects are
small, and the present QCD perturbation calculations are viable and
applicable even in the nuclear physics domain.  The lack of
deviation from the QCD parametrization suggests that the parameter
$\Lambda$ in Eq.~(\ref{eq:14}) is small.  Alternatively, this can be
taken as evidence for fixed point behavior of the QCD coupling in
the infrared. A comparison with a standard definition such as
$\Lambda_{\overline{MS}}$ would require a calculation of
next-to-leading effects.  A more definitive check of QCD can be made
by calculating the normalization of $f_d(Q^2)$ from a perturbative
calculation of  $T_H$ and the evolution of the deuteron wave
function to short distances.  It is also important to confirm
experimentally that the $h=h^\prime=0$ form factor is indeed
dominant.

Note that the deuteron wave function which contributes to the
asymptotic limit of the form factor is the totally antisymmetric
wave function corresponding to the orbital Young symmetry given by
$[6]$ and isospin $(T)+$ spin $(S)$ Young symmetry given by
$\{33\}$. The deuteron state with this symmetry is related to the
$NN$, $\Delta\Delta$, and hidden-color $(CC)$ physical bases, for
both the $(TS)=(01)$ and (10) cases, by the formula
\begin{equation}
\psi_{[6]\{33\}} = \left(\frac{1}{9}\right)^{1/2} \psi_{NN} +
\left(\frac{4}{45}\right)^{1/2} \psi_{\Delta \Delta} +
\left(\frac{4}{5}\right)^{1/2} \psi_{CC} \ .
\label{eq:15}
\end{equation}
Thus the physical deuteron state, which is mostly $\psi_{NN}$ at
large distance, must evolve to the $\psi_{[6]\{33\}}$ state when the
six-quark transverse separations $b^i_\perp \le O(1/Q)\rightarrow
0$.  Since this state is 80\% hidden color, the deuteron wave
functions cannot be described by the nucleonic degrees of freedom in
this domain.  The fact that the six-quark color-singlet state
inevitably evolves in QCD to a dominantly hidden-color configuration
at small transverse separation also has implications for the form of
the nucleon-nucleon potential, which can be considered as one
component in a coupled-channel system.  As the two nucleons approach
each other, the system must do work in order to change the six-quark
state to a dominantly hidden-color configuration; {\em i.e.}, QCD
requires that the nucleon-nucleon potential must be repulsive at
short distances~\cite{Harvey:1980rv}.

Thus a rigorous prediction of QCD is the ``hidden color" of nuclear
wavefunctions at short distances. QCD predicts that nuclear
wavefunctions contain ``hidden
color"~\cite{Matveev:1977xt,Brodsky:1983vf} components: color
configurations not dual to the usual nucleonic degrees of freedom.
In general, the six-quark wavefunction of a deuteron is a mixture of
five different color-singlet states. The dominant color
configuration at large distances corresponds to the usual
proton-neutron bound state where transverse momenta are  of order
${\vec k}^2 \sim 2 M_d \epsilon_{BE}.$ However, at small impact
space separation, all five Fock color-singlet components eventually
acquire equal weight, {\em i.e.}, the deuteron wavefunction evolves
to 80\% hidden color.

\subsection{Hadron Helicity Conservation}

The distribution amplitudes are $L_z=0$ projections of the LF
wavefunction, and the sum of the spin projections of the valence
quarks must equal the $J_z$ of the parent hadron.  Higher orbital
angular momentum components lead to power-law suppressed exclusive
amplitudes~\cite{Lepage:1980fj,Ji:2003fw}.  Since quark masses can
be neglected at leading twist in $T_H$, one has quark helicity
conservation, and thus, finally, hadron-helicity conservation: the
sum of initial hadron helicities equals the sum of final helicities.
In particular, since the hadron-helicity violating Pauli form factor
is computed from states with $\Delta L_z = \pm 1,$  PQCD predicts
$F_2(Q^2)/F_1(Q^2) \sim 1/Q^2 $ [modulo logarithms].  A detailed
analysis shows that the asymptotic fall-off takes the form
$F_2(Q^2)/F_1(Q^2) \sim \log^2 Q^2/Q^2$~\cite{Belitsky:2002kj}.  One
can also construct other models~\cite{Brodsky:2003pw} incorporating
the leading-twist perturbative QCD prediction which are consistent
with the JLab polarization transfer data~\cite{Jones:1999rz} for the
ratio of proton Pauli and Dirac form factors.  This analysis can
also be extended to study the spin structure of scattering
amplitudes at large transverse momentum and other processes which
are dependent on the scaling and orbital angular momentum structure
of light-front wavefunctions.  Recently, Afanasev, Carlson, Chen,
Vanderhaeghen, and I~\cite{Chen:2004tw} have shown that the
interfering two-photon exchange contribution to elastic
electron-proton scattering, including inelastic intermediate states,
can account for the discrepancy between Rosenbluth and Jefferson Lab
spin transfer polarization data~\cite{Jones:1999rz}.

\subsection{Timelike Form Factors}

A crucial prediction of models for proton form factors is the
relative phase of the timelike form factors, since this can be
measured from the proton single spin symmetries in $e^+ e^- \to p
\bar p$ or $p \bar p \to \ell \bar \ell$~\cite{Brodsky:2003gs}. Carl
Carlson, John Hiller, Dae Sung Hwang and I~\cite{Brodsky:2003gs}
have shown that measurements of the proton's polarization strongly
discriminate between the analytic forms of models which fit the
proton form factors in the spacelike region.  In particular, the
single-spin asymmetry normal to the scattering plane measures the
relative phase difference between the timelike $G_E$ and $G_M$ form
factors.  The dependence on proton polarization in the timelike
region is expected to be large in most models, of the order of
several tens of percent.  The continuation of the spacelike form
factors to the timelike domain $t = s > 4 M^2_p$ is very sensitive
to the analytic form of the form factors; in particular it is very
sensitive to the form of the PQCD predictions including the
corrections to conformal scaling.  The forward-backward $\ell^+
\ell^-$ asymmetry can measure the interference of one-photon and
two-photon contributions to $\bar p p \to \ell^+ \ell^-.$

\section{Complications from Final-State Interactions}

Although it has been more than 35 years since the discovery of
Bjorken scaling~\cite{Bjorken:1968dy} in
electroproduction~\cite{Bloom:1969kc}, there are still many issues
in deep-inelastic lepton scattering and Drell-Yan reactions which
are only now being understood from a fundamental basis in QCD.

It is usually assumed---following the parton model---that the
leading-twist structure functions measured in deep  inelastic
lepton-proton scattering are simply the probability distributions
for finding quarks and gluons in the target nucleon.  In fact, gluon
exchange between the fast, outgoing quarks and the target spectators
effects the leading-twist structure functions in a profound way,
leading to  diffractive leptoproduction processes, shadowing of
nuclear structure  functions, and target spin asymmetries.

As I shall discuss in this section, the final-state interactions
from gluon exchange between the outgoing quark and the target
spectator system lead to single-spin asymmetries in semi-inclusive
deep inelastic lepton-proton scattering at leading twist in
perturbative QCD; {\em i.e.}, the rescattering corrections of the
struck quark with the target spectators are not power-law suppressed
at large photon virtuality $Q^2$ at fixed
$x_{bj}$~\cite{Brodsky:2002cx}  The final-state interaction from
gluon exchange occurring immediately after the interaction of the
current also produces a leading-twist diffractive component to deep
inelastic scattering $\ell p \to \ell^\prime p^\prime X$
corresponding to color-singlet exchange with the target system; this
in turn produces shadowing and anti-shadowing of the nuclear
structure functions~\cite{Brodsky:2002ue,Brodsky:1989qz}. In
addition, one can show that the pomeron structure function derived
from diffractive DIS has the same form as the quark contribution of
the gluon structure function~\cite{Brodsky:2004hi}. The final-state
interactions occur at a short light-front time $\Delta\tau \simeq
1/\nu$ after the virtual photon interacts with the struck quark,
producing a nontrivial phase. Here $\nu = p \cdot q/M$ is the
laboratory energy of the virtual photon. Thus none of the above
phenomena is contained in the target light-front wave functions
computed in isolation. In particular, the shadowing of nuclear
structure functions is due to destructive interference effects from
leading-twist diffraction of the virtual photon, physics not
included in the nuclear light-front wave functions.  Thus the
structure functions measured in deep inelastic lepton scattering are
affected by final-state rescattering, modifying their connection to
light-front probability distributions. As an alternative formalism,
one can augment the light-front wave functions with a gauge link
corresponding to an external field created by the virtual photon $q
\bar q$ pair current~\cite{Belitsky:2002sm,Collins:2004nx}.  Such a
gauge link is process dependent~\cite{Collins:2002kn}, so the
resulting augmented LFWFs are not
universal~\cite{Brodsky:2002ue,Belitsky:2002sm,Collins:2003fm}.
Such rescattering corrections are  not contained in the target
light-front wave functions computed in isolation.

Single-spin asymmetries in hadronic reactions provide a remarkable
window to QCD mechanisms at the amplitude level.  In general,
single-spin asymmetries measure the correlation of the spin
projection of a hadron with a production or scattering
plane~\cite{Sivers:1990fh}.  Such correlations are odd under time
reversal, and thus they can arise in a time-reversal invariant
theory only when there is a phase difference between different spin
amplitudes.  Specifically, a nonzero correlation of the proton spin
normal to a production plane measures the phase difference between
two amplitudes coupling the proton target with $J^z_p = \pm {1\over
2}$ to the same final-state.  The calculation requires the overlap
of target light-front wavefunctions with different orbital angular
momentum: $\Delta L^z = 1;$ thus a single-spin asymmetry (SSA)
provides a direct measure of orbital angular momentum in the QCD
bound state.

The shadowing and antishadowing of nuclear structure functions in
the Gribov-Glauber picture is due to the destructive and
constructive coherence, respectively, of amplitudes arising from the
multiple-scattering of quarks in the nucleus.  The effective
quark-nucleon scattering amplitude includes Pomeron and Odderon
contributions from multi-gluon exchange as well as Reggeon quark
exchange contributions~\cite{Brodsky:1989qz}.  The multiscattering
nuclear processes from Pomeron, Odderon and pseudoscalar Reggeon
exchange leads to shadowing and antishadowing of the electromagnetic
nuclear structure functions in agreement with measurements. An
important conclusion is that antishadowing is
non\-universal---different for quarks and antiquarks and different
for strange quarks versus light quarks.  This picture thus leads to
substantially different nuclear effects for charged and neutral
currents, particularly in anti-neutrino reactions, thus affecting
the extraction of the weak-mixing angle $\sin^2\theta_W$ and the
constant $\rho_o$ which are determined from the ratios of charged
and neutral current contributions in deep inelastic neutrino and
anti-neutrino scattering. In recent work, Schmidt, Yang, and
I~\cite{Brodsky:2004qa} have shown that a substantial part of the
difference between the standard model prediction and the anomalous
NuTeV result~\cite{Zeller:2001hh} for $\sin^2\theta_W$ could be due
to the different behavior of nuclear antishadowing for charged and
neutral currents.  Detailed measurements of the nuclear dependence
of charged, neutral and electromagnetic DIS processes are needed to
establish the distinctive phenomenology of shadowing and
antishadowing and to make the NuTeV results definitive.

\subsection{The Paradox of Diffractive Deep Inelastic Scattering}

A remarkable feature of deep inelastic lepton-proton scattering at
HERA is that approximately 10\% events are
diffractive~\cite{Abramowicz:1999eq,Adloff:1997sc,Breitweg:1998gc}:
the target proton remains intact and there is a large rapidity gap
between the proton and the other hadrons in the final state.  These
diffractive deep inelastic scattering (DDIS) events can be
understood most simply from the perspective of the color-dipole
model~\cite{Raufeisen:2000sy}: the $q \bar q$ Fock state of the
high-energy virtual photon diffractively dissociates into a
diffractive dijet system.  The color-singlet exchange of multiple
gluons  between  the color dipole of the $q \bar q$ and the quarks
of the target proton leads to the diffractive final state.  The same
hard pomeron exchange also controls diffractive vector meson
electroproduction at large photon virtuality~\cite{Brodsky:1994kf}.
One can show by analyticity and crossing symmetry that amplitudes
with $C=+$ hard-pomeron exchange have a nearly imaginary phase.

This observation presents a paradox: deep inelastic scattering is
usually discussed in terms of the parton model.  If one chooses the
conventional parton model frame where the photon light-front
momentum is negative $q+ = q^0 + q^z  < 0$, then the virtual photon
cannot produce a virtual $q \bar q$ pair. Instead, the virtual
photon always interacts with a quark constituent with light-front
momentum fraction $x = {k^+\over p^+} = x_{bj}.$  If one chooses
light-front gauge $A^+=0$, then the gauge link associated with the
struck quark (the  Wilson line)  becomes unity. Thus the struck
``current" quark experiences no final-state interactions.  The
light-front wavefunctions $\psi_n(x_i,k_{\perp i}$ of the proton
which determine the quark probability distributions $q(x, Q)$ are
real since the proton is stable.  Thus it appears impossible to
generate the required imaginary phase, let alone the large rapidity
gaps associated with  of DDIS.

This paradox was resolved by Paul Hoyer, Nils Marchal, Stephane
Peigne, Francesco Sannino and myself~\cite{Brodsky:2002ue}.  It is
helpful to consider the case where the virtual photon interacts with
a strange quark --  the $s \bar s$ pair is assumed to be produced in
the target by gluon splitting.  In the case of Feynman gauge, the
struck $s$ quark  continues to interact in the final state via gluon
exchange as described by the Wilson line.  The final-state
interactions occur at a light-front time $\Delta\tau \simeq 1/\nu$
after the virtual photon interacts with the struck quark.  When one
integrates over the nearly-on-shell intermediate state, the
amplitude acquires an imaginary part.  Thus the rescattering of the
quark produces a separated color-singlet $s \bar s$ and an imaginary
phase.

In contrast, in the case of the light-front gauge $A^+ = n \cdot A
=0$, one must consider the final state interactions of the
(unstruck) $\bar s$ quark.  light-front gauge  is singular---in
particular, the gluon propagator \begin{equation} d_{LC}^{\mu\nu}(k)
= \frac{i}{k^2+\ieps}\left[-g^{\mu\nu}+\frac{n^\mu k^\nu+ k^\mu
n^\nu}{n\cdot k}\right] \label{lcprop} \end{equation} has a pole at
$k^+ = 0$ which requires an analytic prescription.  In final-state
scattering involving nearly on-shell intermediate states, the
exchanged momentum $k^+$ is of \order{1/\nu} in the target rest
frame, which enhances the second term in the propagator.  This
enhancement allows rescattering to contribute at leading twist even
in LC gauge.  Thus the  rescattering contribution survives in the
Bjorken limit because of the singular behavior of the propagator of
the exchanged gluon at small $k^+$ in $A^+=0$ gauge.    The net
result is  gauge invariant and identical to the color dipole model
calculation.

\vspace{0.3cm}
\begin{figure}[htb]
\begin{center}
\includegraphics[width=4.5in]{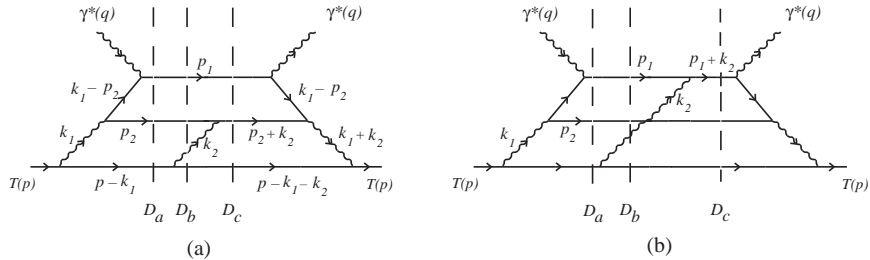}
\end{center}
\caption[*]{\baselineskip 13pt Two types of final state
interactions.  (a) Scattering of the antiquark ($p_2$ line), which
in the aligned jet kinematics is part of the target dynamics.  (b)
Scattering of the current quark ($p_1$ line).  For each light-front
time-ordered diagram, the potentially on-shell intermediate
states---corresponding to the zeroes of the denominators $D_a, D_b,
D_c$---are denoted by dashed lines.} \label{brodsky1}
\end{figure}

The calculation of the rescattering effects on DIS in Feynman and
light-front gauge through three loops is given in detail for a
simple Abelian model in Ref.~\cite{Brodsky:2002ue}.
Figure~\ref{brodsky1} illustrates two LCPTH diagrams which
contribute to the forward $\gamma^* T \to \gamma^* T$ amplitude,
where the target $T$ is taken to be a single quark. In the aligned
jet kinematics the virtual photon fluctuates into a \qu\qb\ pair
with limited transverse momentum, and the (struck) quark takes
nearly all the longitudinal momentum of the photon. The initial \qu\
and \qb\ momenta are denoted $p_1$ and $p_2-k_1$, respectively.  The
result is most easily expressed in eikonal form in terms of
transverse distances $r_T, R_T$ conjugate to $p_{2T}, k_T$.  The DIS
cross section can be expressed as
\begin{equation}
Q^4\frac{d\sigma}{dQ^2\, dx_B} = \frac{\alpha_{\rm
em}}{16\pi^2}\frac{1-y}{y^2} \frac{1}{2M\nu} \int
\frac{dp_2^-}{p_2^-}\,d^2\rvec_T\, d^2\Rvec_T\, |\tilde M|^2
\label{transcross} \end{equation} where \begin{equation} |\tilde{
M}(p_2^-,\rvec_T, \Rvec_T)| = \left|\frac{\sin \left[g^2\,
W(\rvec_T, \Rvec_T)/2\right]}{g^2\, W(\rvec_T, \Rvec_T)/2}
\tilde{A}(p_2^-,\rvec_T, \Rvec_T)\right| \label{Interference}
\end{equation}
is the resummed result.  The Born amplitude is
\begin{equation} \tilde
A(p_2^-,\rvec_T, \Rvec_T) = 2eg^2 M Q p_2^-\, V(m_\pl r_T)
W(\rvec_T, \Rvec_T) \label{Atildeexpr} \end{equation}
where $m_\pl^2 = p_2^-Mx_B + m^2 \label{mplus}$ and
\begin{equation}
V(m\, r_T) \equiv \int \frac{d^2\pvec_T}{(2\pi)^2}
\frac{e^{i\rvec_T\cdot\pvec_{T}}}{p_T^2+m^2} =
\frac{1}{2\pi}K_0(m\,r_T). \label{Vexpr} \end{equation}
The
rescattering effect of the dipole of the \qu\qb~ is controlled by
\begin{equation} W(\rvec_T, \Rvec_T) \equiv \int \frac{d^2\kvec_T}{(2\pi)^2}
\frac{1-e^{i\rvec_T\cdot\kvec_{T}}}{k_T^2}
e^{i\Rvec_T\cdot\kvec_{T}} = \frac{1}{2\pi}
\log\left(\frac{|\Rvec_T+\rvec_T|}{R_T} \right). \label{Wexpr}
\end{equation}
The fact that the coefficient of $\tilde A$ in is less than unity
for all $\rvec_T, \Rvec_T$ shows that the rescattering corrections
reduce the cross section in  analogy  to nuclear shadowing.

A new understanding of the role of final-state interactions in deep
inelastic scattering has thus emerged.  The final-state interactions
from gluon exchange occurring immediately after the interaction of
the current produce a leading-twist diffractive component to deep
inelastic scattering $\ell p \to \ell^\prime p^\prime X$ due to the
color-singlet exchange with the target system.  This rescattering is
described in the Feynman gauge by the path-ordered exponential
(Wilson line) in the expression for the parton distribution function
of the target.  The multiple  scattering of the struck parton via
instantaneous interactions in the target generates dominantly
imaginary diffractive amplitudes, giving rise to an effective ``hard
pomeron" exchange.  The presence of a rapidity gap between the
target and diffractive system requires that the target remnant
emerges in a color-singlet state; this is made possible in any gauge
by the soft rescattering of the final-state $s -\bar s$ system.

\subsection{Diffractive Deep Inelastic Reactions and Rescattering}

Rikard Enberg, Paul Hoyer, Gunnar Ingelman and I have recently
discussed some further aspects of the QCD dynamics of diffractive
deep inelastic scattering~\cite{Brodsky:2004hi}. We show that the
quark structure function of the effective hard pomeron has the same
form as the quark contribution of the gluon structure function. The
hard pomeron  is not an intrinsic part of the proton; rather it must
be considered as  a dynamical effect of the lepton-proton
interaction.

Our QCD-based picture also applies to diffraction in
hadron-initiated processes.  The rescattering is different in
virtual photon- and hadron-induced processes due to the different
color environment, which accounts for the  observed non-universality
of diffractive parton distributions.  In the hadronic case the color
flow at tree level can involve color-octet as well as color-triplet
separation.  Multiple scattering of the quarks and gluons can set up
a variety of different color singlet domains.  This framework also
provides a theoretical basis for the phenomenologically successful
Soft Color Interaction (SCI) model which includes rescattering
effects and thus generates a variety of final states  with rapidity
gaps.

\subsection{Origin of Nuclear Shadowing and Antishadowing}

The physics of nuclear shadowing in deep inelastic scattering can be
most easily understood in the laboratory frame using the
Glauber-Gribov
picture~\cite{Glauber:1955qq,Gribov:1968gs,Gribov:1968jf}.  The
virtual photon, $W,$ or $Z^0$  produces a quark-antiquark
color-dipole pair which can interact diffractively or inelastically
on the nucleons in the nucleus.  The destructive interference of
diffractive amplitudes from pomeron exchange on the upstream
nucleons then causes shadowing of the virtual photon interactions on
the back-face
nucleons~\cite{Stodolsky:1966am,Brodsky:1969iz,Brodsky:1990qz,Ioffe:1969kf,
Frankfurt:1988zg,Kopeliovich:1998gv,Kharzeev:2002fm}. The
Bjorken-scaling diffractive interactions on the nucleons in a
nucleus thus leads to the shadowing (depletion at small $x_{bj}$) of
the nuclear structure functions.

\begin{figure}[htb]
\begin{center}
\includegraphics[height=3in]{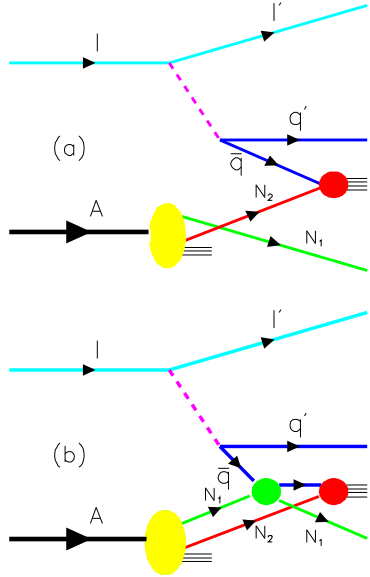}
\end{center}
\caption[*]{\baselineskip 13pt The one-step and two-step processes
in DIS on a nucleus.  If the scattering on nucleon $N_1$ is via
pomeron exchange, the one-step and two-step amplitudes are
opposite in phase, thus diminishing the $\bar q$ flux reaching
$N_2.$ This causes shadowing of the charged and neutral current
nuclear structure functions. \label{bsy1f1}}
\end{figure}

As emphasized by Ioffe~\cite{Ioffe:1969kf}, the coherence between
processes which occur on different nucleons at separation $L_A$
requires small Bjorken $x_{B}:$ $1/M x_B = {2\nu/ Q^2}  \ge L_A .$
The coherence between different quark processes is also the basis of
saturation phenomena in DIS and other hard QCD reactions at small
$x_B$~\cite{Mueller:2004se}, and coherent multiple parton scattering
has been used in the analysis of $p+A$ collisions in terms of the
perturbative QCD factorization approach~\cite{Qiu:2004da}.  An
example of the interference of one- and two-step processes in deep
inelastic lepton-nucleus scattering illustrated in
Fig.~\ref{bsy1f1}.

\vspace{0.3cm}

An important aspect of the shadowing phenomenon is that the
diffractive contribution $\gamma^* N \to X N^\prime$ to deep
inelastic scattering (DDIS) where the nucleon $N_1$ in
Fig.~\ref{bsy1f1} remains intact is a constant fraction of the total
DIS rate, confirming that it is a leading-twist contribution.  The
Bjorken scaling of DDIS has been observed at
HERA~\cite{Adloff:1997sc,Martin:2004xw,Ruspa:2004jb}.  As shown in
Ref.~\cite{Brodsky:2002ue}, the leading-twist contribution to DDIS
arises in QCD in the usual parton model frame when one includes the
nearly instantaneous gluon exchange final-state interactions of the
struck quark with the target spectators.  The same final state
interactions also lead to leading-twist single-spin asymmetries in
semi-inclusive DIS~\cite{Brodsky:2002cx}.  Thus the shadowing of
nuclear structure functions is also a leading-twist effect.

It was shown in Ref.~\cite{Brodsky:1989qz}  that if one allows for
Reggeon exchanges which leave a nucleon intact,  then one can obtain
{\it constructive} interference among the multi-scattering
amplitudes in the nucleus.   A Bjorken-scaling contribution to DDIS
from Reggeon exchange has in fact also been observed at
HERA~\cite{Adloff:1997sc,Ruspa:2004jb}.  The strength and energy
dependence of the $C=+$ Reggeon $t-$channel exchange contributions
to virtual Compton scattering is constrained by the
Kuti-Weisskopf~\cite{Kuti:1971ph} behavior $F_2(x) \sim
x^{1-\alpha_R}$ of the non-singlet electromagnetic structure
functions at small $x$.  The phase of the Reggeon exchange amplitude
is determined by its signature factor.  Because of this phase
structure~\cite{Brodsky:1989qz}, one obtains constructive
interference and {\it antishadowing} of the nuclear structure
functions in the range $0.1 < x < 0.2$ -- a pronounced excess of the
nuclear cross section with respect to nucleon
additivity~\cite{Arneodo:1992wf}.

In the case where the diffractive amplitude on $N_1$ is imaginary,
the two-step process has the phase $i \times i = -1 $ relative to
the one-step amplitude, producing destructive interference.  (The
second factor of $i$ arises from integration over the quasi-real
intermediate state.)  In the case where the diffractive amplitude on
$N_1$ is due to $C=+$ Reggeon exchange with intercept $\alpha_R(0) =
1/2$, for example, the phase of the two-step amplitude is ${1\over
\sqrt 2}(1-i) \times i = {1\over \sqrt 2} (i+1)$ relative to the
one-step amplitude, thus producing constructive interference and
antishadowing.

The effective quark-nucleon scattering amplitude includes Pomeron
and Odderon contributions from multi-gluon exchange as well as
Reggeon quark-exchange contributions~\cite{Brodsky:1989qz}.  The
coherence of these multiscattering nuclear processes leads to
shadowing and antishadowing of the electromagnetic nuclear structure
functions in agreement with measurements.  The Reggeon contributions
to the quark scattering amplitudes depend specifically on the quark
flavor; for example the isovector Regge trajectories couple
differently to $u$ and $d$ quarks.  The $s$ and $\bar s$ couple to
yet different Reggeons.  This implies distinct anti-shadowing
effects for each quark and antiquark component of the nuclear
structure function. Ivan Schmidt,  Jian-Jun Yang, and
I~\cite{Brodsky:2004bg} have shown that this picture leads to
substantially different antishadowing for charged and neutral
current reactions.

\begin{figure}[htb]
\begin{center}
\includegraphics[width=4in,height=3.5in]{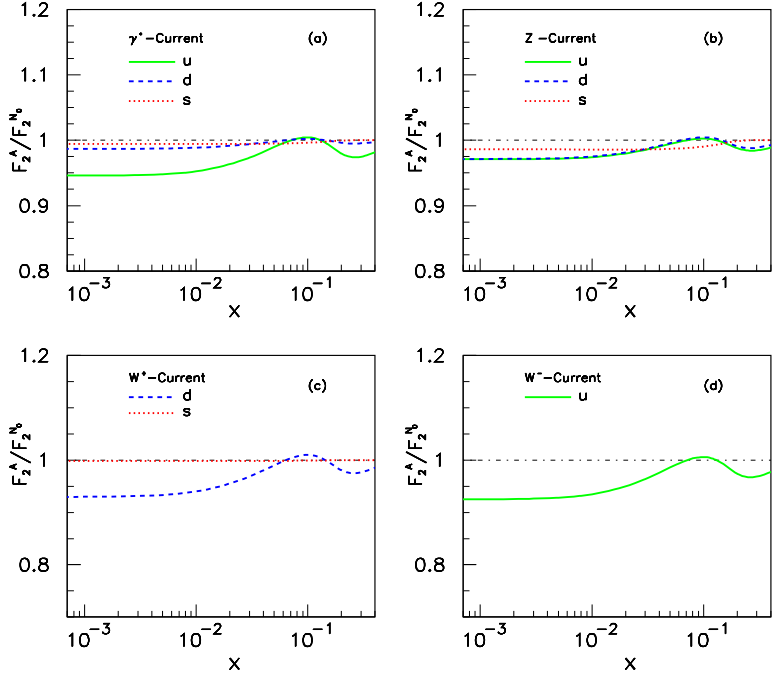}
\end{center}
\caption[*]{\baselineskip 13pt The  quark contributions to the
ratios of structure functions at $ Q^2 = 1~\rm{GeV}^2$.  The solid,
dashed and dotted curves correspond to the $u$, $d$ and $s$ quark
contributions, respectively.  This corresponds in our model to the
nuclear dependence of the $\sigma(\bar u-A)$, $\sigma(\bar d-A)$,
$\sigma(\bar s-A)$ cross sections, respectively.  In order to stress
the individual contribution of quarks, the numerator of the ratio
$F_2^{A} / F_2^{N_0}$ shown in these two figures is obtained from
the denominator by a replacement $q^{N_0}$ into $q^{A}$ for only the
considered quark.  As a result, the effect of antishadowing appears
diminished.
 \label{bsy1f5}}
\end{figure}

\begin{figure}[ht]
\begin{center}
\includegraphics[width=4in,height=3.5in]{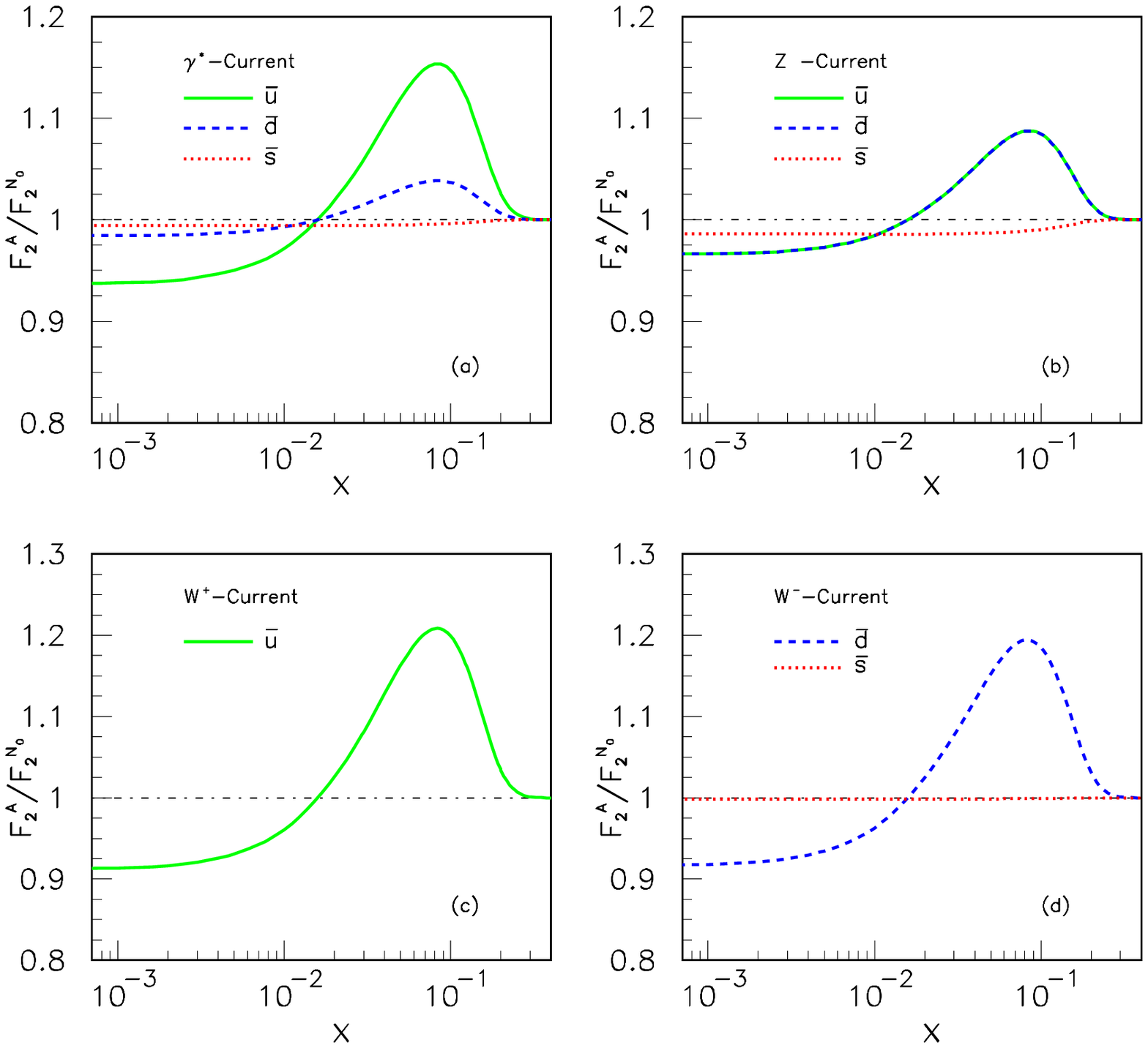} 
\end{center}
\caption[*]{\baselineskip 13pt The anti-quark contributions to
ratios of the structure functions at $ Q^2 = 1~\rm{GeV}^2$.  The
solid, dashed and dotted curves correspond to $\bar{u}$, $\bar{d}$
and $\bar{s}$ quark contributions, respectively.  This corresponds
in our model to the nuclear dependence of the $\sigma(u-A)$,
$\sigma(d-A)$, $\sigma(s-A)$ cross sections, respectively.  In order
to stress the individual contribution of quarks, the numerator of
the ratio $F_2^{A} / F_2^{N_0}$ shown in these two figures is
obtained from the denominator by a replacement $\bar{q}^{N_0}$ into
$\bar{q}^{A}$ for only the considered anti-quark.
 \label{bsy1f6}}
\end{figure}

Figures~\ref{bsy1f5}--\ref{bsy1f6} illustrate the individual quark
$q$ and anti-quark $\bar{q}$ contributions to the ratio of the iron
to nucleon structure functions $R=F_2^{A} / F_2^{N_0}$ in a model
calculation where the Reggeon contributions are constrained by the
Kuti-Weisskopf behavior~\cite{Kuti:1971ph} of the nucleon structure
functions at small $x_{bj}.$  Because the strange quark distribution
is much smaller than $u$ and $d$ quark distributions, the strange
quark contribution to the ratio is very close to 1 although
$s^{A}/s^{N_0}$ may significantly deviate from 1.

\vspace{0.3cm}

\vspace{0.3cm}

Our analysis leads to substantially different nuclear antishadowing
for charged and neutral current reactions; in fact, the neutrino and
antineutrino DIS cross sections are each modified in different ways
due to the various allowed Regge exchanges.  The non-universality of
nuclear effects will modify the extraction of the weak-mixing angle
$\sin^2\theta_W$, particularly because of the strong nuclear effects
for the $F_3$ structure function.  The shadowing and antishadowing
of the strange quark structure function in the nucleus can also be
considerably different than that of the light quarks. We thus find
that part of the anomalous NuTeV result~\cite{McFarland:2003gx} for
$\sin^2\theta_W$ could be due to the non-universality  of nuclear
antishadowing for charged and neutral currents.  Our picture also
implies non-universality for the nuclear modifications of
spin-dependent structure functions.

Thus the antishadowing of nuclear structure functions depends in
detail on quark flavor. Careful measurements of the nuclear
dependence of charged, neutral, and electromagnetic DIS processes
are needed to establish the distinctive phenomenology of shadowing
and antishadowing and to make the NuTeV results definitive.  It is
also important to map out the shadowing and antishadowing of each
quark component of the nuclear structure functions to illuminate the
underlying QCD mechanisms.  Such studies can be carried out in
semi-inclusive deep inelastic scattering for the electromagnetic
current at Hermes and at Jefferson Laboratory by tagging the flavor
of the current quark or by using pion and kaon-induced Drell-Yan
reactions.  A new determination of $\sin^2\theta_W$ is also expected
from the neutrino scattering experiment NOMAD at CERN~\cite{Petti}.
A systematic program of measurements of the nuclear effects in
charged and neutral current reactions could also be carried out in
high energy electron-nucleus colliders such as HERA and eRHIC, or by
using high intensity neutrino beams~\cite{Geer}.

\subsection{ Single-Spin Asymmetries from Final-State Interactions}

Spin correlations provide a remarkably sensitive window to
hadronic structure and basic mechanisms in QCD.   Among the most
interesting polarization effects are single-spin azimuthal
asymmetries  in semi-inclusive deep inelastic scattering,
representing the correlation of the spin of the proton target and
the virtual photon to hadron production plane: $\vec S_p \cdot
\vec q \times \vec p_H$~\cite{Avakian:2002td}.  Such asymmetries
are time-reversal odd, but they can arise in QCD through phase
differences in different spin amplitudes.

Until recently, the traditional explanation of pion
electroproduction single-spin asymmetries in semi-inclusive deep
inelastic scattering is that they are proportional to the
transversity distribution of the quarks in the hadron
$h_{1}$~\cite{Jaffe:1996zw,Boer:2001zw,Boer:2002xc} convoluted with
the transverse momentum dependent fragmentation (Collins) function
$H^\perp_1$, the distribution for a transversely polarized quark to
fragment into an unpolarized hadron with non-zero transverse
momentum
\cite{Collins93,Barone:2001sp,Ma:2002ns,Goldstein:2002vv,Gamberg:2003ey}.

\begin{figure}[htb]
\begin{center}
\includegraphics[width=3in,height=3in]{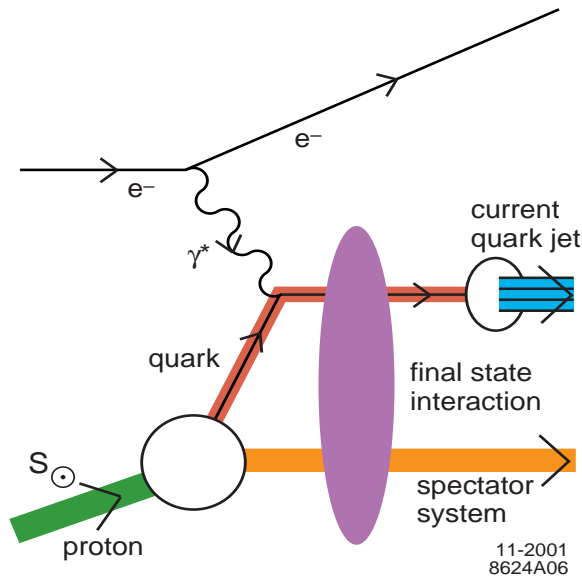}
\end{center}
\caption[*]{\baselineskip 13pt The origin of the Sivers effect in
semi-inclusive deep inelastic scattering} \label{SSA}
\end{figure}

Dae Sung Hwang, Ivan Schmidt and I have showed that an alternative
physical mechanism for the azimuthal asymmetries  also
exists~\cite{Brodsky:2002cx,Collins,Ji:2002aa}. The same QCD
final-state interactions (gluon exchange) between the struck quark
and the proton spectators  which leads to diffractive events also
can produce single-spin asymmetries (the Sivers effect) in
semi-inclusive deep inelastic lepton scattering which survive in the
Bjorken limit.  This is illustrated in Fig.~\ref{SSA}.  In contrast
to the SSAs arising from transversity and the Collins fragmentation
function, the fragmentation of the quark into hadrons is not
necessary;  one predicts a correlation with the production plane of
the quark jet itself $\vec S_p \cdot \vec q \times \vec p_q.$

\vspace{0.3cm}

The final-state interaction mechanism provides an appealing physical
explanation within QCD of single-spin asymmetries.  Remarkably, the
same matrix element which determines the spin-orbit correlation
$\vec S \cdot \vec L$  also produces the anomalous magnetic moment
of the proton, the Pauli form factor, and the generalized parton
distribution $E$ which is measured in deeply virtual Compton
scattering.  Physically, the final-state interaction phase arises as
the infrared-finite difference of QCD Coulomb phases for hadron wave
functions with differing orbital angular momentum.  An elegant
discussion of the Sivers effect including its sign has been given by
Burkardt~\cite{Burkardt:2004vm}.

The final-state interaction effects can also be identified with the
gauge link which is present in the gauge-invariant definition of
parton distributions~\cite{Collins}.  Even when the light-front
gauge is chosen, a transverse gauge link is required.  Thus in any
gauge the parton amplitudes need to be augmented by an additional
eikonal factor incorporating the final-state interaction and its
phase~\cite{Ji:2002aa,Belitsky:2002sm}. The net effect is that it is
possible to define transverse momentum dependent parton distribution
functions which contain the effect of the QCD final-state
interactions.

A related analysis also predicts that the initial-state
interactions from gluon exchange between the incoming quark and
the target spectator system lead to leading-twist single-spin
asymmetries in the Drell-Yan process $H_1 H_2^\updownarrow \to
\ell^+ \ell^- X$ \cite{Collins:2002kn,BHS2}.    Initial-state
interactions also lead to a $\cos 2 \phi$ planar correlation in
unpolarized Drell-Yan reactions \cite{Boer:2002ju}.

\subsection{Calculations of Single-Spin Asymmetries in QCD}

Hwang, Schmidt and I have calculated \cite{Brodsky:2002cx} the
single-spin Sivers asymmetry in semi-inclusive electroproduction
$\gamma^* p^{\updownarrow} \to H X$ induced by final-state
interactions in a model of a spin-1/2 ~ proton of mass $M$ with
charged spin-1/2~ and spin-0 constituents of mass $m$ and $\lambda$,
respectively, as in the QCD-motivated quark-diquark model of a
nucleon.  The basic electroproduction reaction is then $\gamma^* p
\to q (qq)_0$.  In fact, the asymmetry comes from the interference
of two amplitudes which have different proton spin, but couple to
the same final quark spin state, and therefore it involves the
interference of tree and one-loop diagrams with a final-state
interaction.  In this simple model the azimuthal target single-spin
asymmetry $A^{\sin \phi}_{UT}$ is given by
\begin{eqnarray}
A^{\sin \phi}_{UT} &=& {C_F \alpha_s(\mu^2) } \ { \Bigl(\ \Delta\,
M+m\ \Bigr)\ r_{\perp}\over \Big[\ \Bigl( \ \Delta\, M+m\
\Bigr)^2\
+\ {\vec r}_{\perp}^2\ \Big]}\nonumber \\
&\times& \Bigg[\ {\vec r}_{\perp}^2+\Delta
(1-\Delta)(-M^2+{m^2\over\Delta} +{\lambda^2\over 1-\Delta})\
\Bigg] \nonumber\\[1ex] &\times& \ {1\over {\vec r}_{\perp}^2}\
{\rm ln}{{\vec r}_{\perp}^2 +\Delta
(1-\Delta)(-M^2+{m^2\over\Delta}+{\lambda^2\over 1-\Delta})\over
\Delta (1-\Delta)(-M^2+{m^2\over\Delta}+{\lambda^2\over
1-\Delta})}\ . \label{sa2b}
\end{eqnarray}
Here $r_\perp$ is the magnitude of the transverse momentum of the
current quark jet relative to the virtual photon direction, and
$\Delta=x_{Bj}$ is the usual Bjorken variable.  To obtain
(\ref{sa2b}) from Eq. (21) of \cite{Brodsky:2002cx}, we used the
correspondence ${|e_1 e_2|/ 4 \pi} \to C_F \alpha_s(\mu^2)$
and the fact that the sign of the charges $e_1$ and $e_2$ of the
quark and diquark are opposite since they constitute a bound
state.  The result can be tested in jet production using an
observable such as thrust to define the  momentum $q + r$ of the
struck quark.

\vspace{0.3cm}
\begin{figure}[htp]
\begin{center}
\includegraphics[width=4in,height=6in]{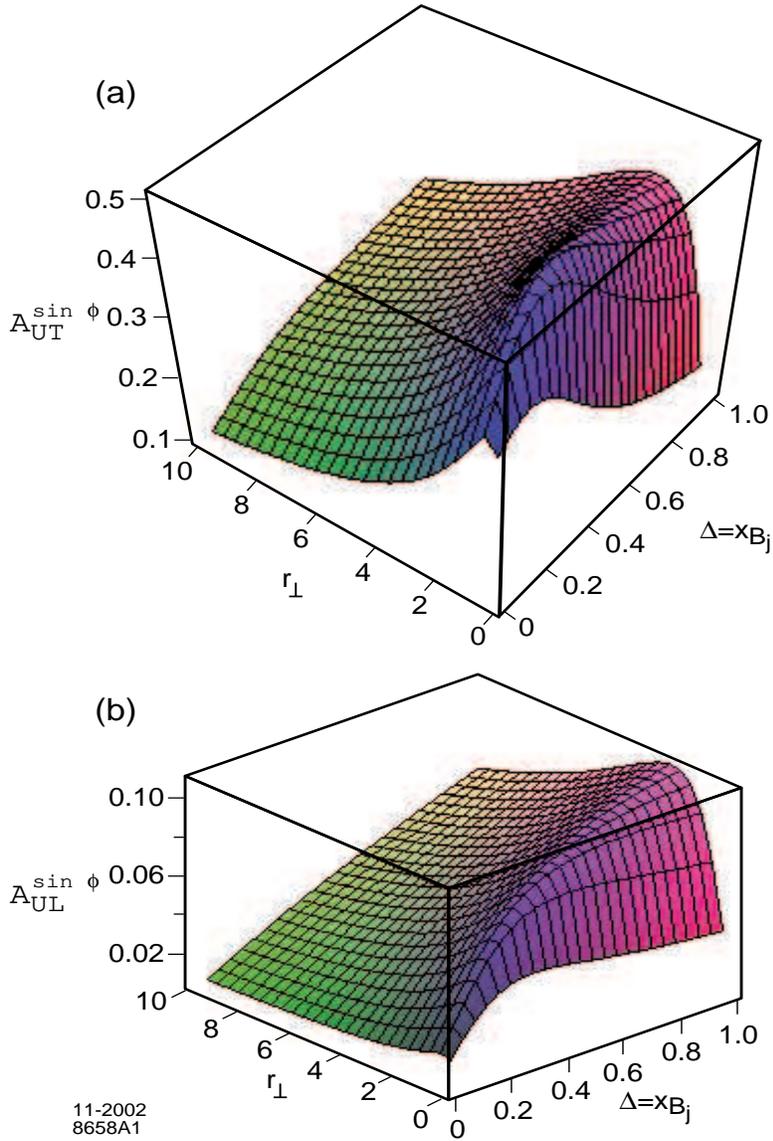}
\end{center}
\caption[*]{\baselineskip 13pt
 Model predictions for the target single-spin asymmetry
$A^{\sin \phi}_{UT}$ for charged and neutral current deep inelastic
scattering resulting from gluon exchange in the final state.  Here
$r_\perp$ is the magnitude of the transverse momentum of the
outgoing quark relative to the photon or vector boson direction, and
$\Delta = x_{bj}$ is the light-front momentum fraction of the struck
quark. The parameters of the model are given in the text.  In (a)
the target polarization is transverse to the incident lepton
direction.  The asymmetry in (b) $A^{\sin \phi}_{UL} = K A^{\sin
\phi}_{UT}$ includes a kinematic factor $K = {Q\over \nu}\sqrt{1-y}$
for the case where the target nucleon is polarized along the
incident lepton direction.  For illustration, we have taken $K= 0.26
\sqrt x,$ corresponding to the kinematics of the HERMES
experiment~\cite{Airapetian:1999tv} with $E_{lab} = 27.6 ~{\rm GeV}$
and $y = 0.5.$} \label{fig:SSA1}
\end{figure}

The predictions of our model for the asymmetry $A^{\sin
\phi}_{UT}$ of the  ${\vec S}_p \cdot \vec q \times \vec p_q$
correlation based on  Eq. ({\ref{sa2b}}) are shown in Fig.
\ref{fig:SSA1}.  As representative parameters we take $\alpha_s =
0.3$, $M =  0.94$ GeV for the proton mass,  $m=0.3$ GeV for the
fermion constituent and $\lambda = 0.8$ GeV for the spin-0
spectator.  The single-spin asymmetry $A^{\sin \phi}_{UT}$ is
shown as a function of $\Delta$ and $r_\perp$ (GeV).  The
asymmetry measured at HERMES~\cite{Airapetian:1999tv}
$A_{UL}^{\sin \phi} = K A^{\sin \phi}_{UT}$ contains a kinematic
factor $K = {Q\over \nu}\sqrt{1-y} = {\sqrt{2Mx\over
E}}{\sqrt{1-y\over y}}$ because the proton is polarized along the
incident electron direction.  The resulting prediction for
$A_{UL}^{\sin \phi}$ is shown in Fig. \ref{fig:SSA1}(b).  Note
that $\vec r = \vec p_q - \vec q$ is the momentum of the current
quark jet relative to the photon momentum.  The asymmetry as a
function of the pion momentum $\vec p_\pi$ requires a convolution
with the quark fragmentation function.

Since the same matrix element controls the Pauli form factor, the
contribution of each quark current to the SSA is proportional to the
contribution $\kappa_{q/p}$ of that quark to the proton target's
anomalous magnetic moment $\kappa_p = \sum_q e_q
\kappa_{q/p}$~\cite{Brodsky:2002cx,Burkardt:2004vm}.
Avakian~\cite{Avakian:2002td} has shown that the data from HERMES
and Jefferson laboratory could be accounted for by the above
analysis. The HERMES collaboration has recently measured the SSA in
pion electroproduction using transverse target
polarization~\cite{Airapetian:2004tw}. The Sivers and Collins
effects can be separated using planar correlations; both
contributions are observed to contribute, with values not in
disagreement with theory expectations.

It should be emphasized that the Sivers effect occurs even for jet
production; unlike transversity, hadronization is not required.
There is no Sivers effect in charged current reactions since the $W$
only couples to left-handed quarks~\cite{Brodsky:2002pr}.

The corresponding single spin asymmetry for the Drell-Yan processes,
such as \break $\pi p^{\leftrightarrow}\ ({\rm or}\ p
p^{\leftrightarrow}) \to \gamma^* X\to \ell^+\ell^- X$, is due to
initial-state interactions.  The simplest way to get the result is
applying crossing symmetry to the SIDIS processes.  The result that
the SSA in the Drell-Yan process is the same as that obtained in
SIDIS, with the appropriate identification of variables, but with
the opposite sign~\cite{Collins,BHS2}.

We can also consider the SSA of $e^+e^-$ annihilation processes such
as $e^+e^-\to \gamma^* \to \pi {\Lambda}^{\leftrightarrow} X$.  The
$\Lambda$ reveals its polarization via its decay $\Lambda \to p
\pi^-$.  The spin of the $\Lambda$ is normal to the decay plane.
Thus we can look for a SSA through the T-odd correlation
$\epsilon_{\mu \nu \rho \sigma} S^\mu_\Lambda p^\nu_\Lambda
q^\rho_{\gamma^*} p^\sigma_{\pi}$. This is related by crossing to
SIDIS on a $\Lambda$ target.

Measurements from Jefferson Lab~\cite{Avakian:2003pk} also show
significant beam single spin asymmetries in deep inelastic
scattering.  Afanasev and Carlson~\cite{Afanasev:2003ze} have
recently shown that this asymmetry is due to the interference of
longitudinal and transverse photoabsorption amplitudes which have
different phases induced by the final-state interaction between
the struck quark and the target spectators just as in the
calculations of Ref. \cite{Brodsky:2002cx}.  Their results are
consistent with the experimentally observed magnitude of this
effect.  Thus similar FSI mechanisms involving quark orbital
angular momentum appear to be responsible for both target and beam
single-spin asymmetries.

\section{New Directions for QCD}

As I have emphasized in these lectures, the light-front
wavefunctions of hadrons are the central elements of QCD
phenomenology, describing bound states in terms of their fundamental
quark and gluon degrees of freedom at the amplitude level.  Given
the light-front wavefunctions one can compute quark and gluon
distributions, distribution amplitudes, generalized parton
distributions, form factors, and matrix elements of local currents
such as semileptonic $B$ decays. The diffractive dissociation of
hadrons on nucleons or nuclei into jets or leading hadrons can
provide new measures of the LFWFs of the projectile as well as tests
of color transparency, hidden color,  and intrinsic charm. The
advent of the $12$ GeV upgrade of the Jefferson Laboratory electron
accelerator and the new $15$ GeV antiproton storage ring HESR at GSI
will open up important new tests of these properties of QCD in
hadronic and nuclear reactions.

Although we are still far from solving QCD explicitly, a number of
properties of the light-front wavefunctions of the hadrons are known
from both phenomenology and the basic properties of QCD.  For
example, the endpoint behavior of light-front wavefunctions and
structure functions can be determined from perturbative arguments
and Regge arguments.   There are also correspondence principles.
For example, for heavy quarks in the nonrelativistic limit, the
light-front formalism reduces to conventional many-body
Schr\"odinger theory.  On the other hand, one can also build
effective three-quark models which encode the static properties of
relativistic baryons.

It is thus imperative to compute the light-front wavefunctions from
first principles in QCD.  Lattice gauge theory can provide moments
of the distribution amplitudes by evaluating  vacuum-to-hadron
matrix elements of local operators~\cite{DelDebbio:1999mq}.  The
transverse lattice is also providing new nonperturbative
information~\cite{Dalley:2004rq,Burkardt:2001jg}. The DLCQ method is
also a first-principles method for solving nonperturbative QCD; at
finite harmonic resolution $K$ the DLCQ Hamiltonian acts in physical
Minkowski space as a finite-dimensional Hermitian matrix in Fock
space.  The DLCQ Heisenberg equation is Lorentz-frame independent
and has the advantage of providing not only the spectrum of hadrons,
but also the complete set of LFWFs for each hadron eigenstate. An
important feature the light-front formalism is that $J_z$ is
conserved; thus one simplify the DLCQ method by projecting the full
Fock space on states with specific angular momentum. As shown in
Ref. ~\cite{Brodsky:2003pw}, the Karmanov-Smirnov operator uniquely
specifies the form of the angular dependence of the light-front
wavefunctions, allowing one to transform the light-front Hamiltonian
equations to  differential equations acting on scalar forms. A
complementary method would be to construct the $T$-matrix for
asymptotic $q \bar q$ or $qqq$ or gluonium states using the
light-front analog of the Lippmann-Schwinger method.  This allows
one to focus on states with the specific global quantum numbers and
spin of a given hadron. The zeros of the resulting resolvent then
provides the hadron spectrum and the respective light-front Fock
state projections.

In principle, the complete spectrum and bound-state wave functions
of a quantum field theory can be determined by finding the
eigenvalues and eigensolutions of its light-cone Hamiltonian.

The DLCQ method has a number of attractive features for solving 3+1
quantum field theories nonperturbatively because of the ability to
truncate the Fock state to low particle number sectors. One of the
challenges in obtaining nonperturbative solutions for gauge theories
such as QCD using light-cone Hamiltonian methods is to renormalize
the theory while preserving Lorentz symmetries and gauge invariance.
For example, the truncation of the light-cone Fock space leads to
uncompensated ultraviolet divergences.  Recently we presented two
methods for consistently regularizing light-cone-quantized gauge
theories in Feynman and light-cone gauges~\cite{Brodsky:2004cx}: (1)
the introduction of a spectrum of Pauli-Villars fields which
produces a finite theory while preserving Lorentz invariance; (2)
the augmentation of the gauge-theory Lagrangian with higher
derivatives.  Finite-mass Pauli-Villars regulators can also be used
to compensate for neglected higher Fock states. As a test case, we
have applied these regularization procedures to an approximate
nonperturbative computation of the anomalous magnetic moment of the
electron in QED as a first attempt to meet Feynman's famous
challenge.

\subsection{Testing Hidden Color}

In traditional nuclear physics, the deuteron is a bound state of a
proton and a neutron where the binding force arise from the exchange
of a pion and other mesonic states. However,  as I have reviewed,
QCD provides a new
perspective~\cite{Brodsky:1976rz,Matveev:1977xt}.: 6 quarks in the
fundamental $3_C$ representation of $SU(3)$ color can combine into 5
different color-singlet combinations, only one of which corresponds
to a proton and neutron In fact, if the deuteron wavefunction is a
proton-neutron bound state at large distances, then as their
separation becomes smaller, the QCD evolution resulting from colored
gluon exchange introduce 4 other ``hidden color" states into the
deuteron wavefunction~\cite{Brodsky:1983vf}. As I have discussed,
the normalization of the deuteron form factor observed at large
$Q^2$~\cite{Arnold:1975dd}, as well as the presence of two mass
scales in the scaling behavior of the reduced deuteron form
factor~\cite{Brodsky:1976rz} thus suggests sizable hidden-color Fock
state contributions such as $\ket{(uud)_{8_C} (ddu)_{8_C}}$ with
probability  of order $15\%$ in the deuteron
wavefunction~\cite{Farrar:1991qi}.

The hidden color states of the deuteron can be materialized at the hadron
level as $\Delta^{++}(uuu) \Delta^{-}(ddd)$ and other novel quantum
fluctuations of the deuteron.  These dual hadron components become more and
more important as one probes the deuteron at short distances, such as in
exclusive reactions at large momentum transfer.  For example, the ratio
$${{d \sigma\over dt}(\gamma d \to
\Delta^{++} \Delta^{-})\over {d\sigma\over dt}(\gamma d\to n p) }$$
should increase dramatically with increasing transverse momentum $p_T.$
Similarly the Coulomb dissociation of the deuteron into various exclusive
channels
$$e d \to e^\prime + p n, p p \pi^-, \Delta \Delta, \cdots$$ should have a
changing composition as the final-state hadrons are probed at high
transverse momentum, reflecting the onset of hidden color degrees of
freedom.

\subsection{Perspectives on QCD from AdS/CFT}

An outstanding  consequence of Maldacena's
duality~\cite{Maldacena:1997re} between 10-dimensional string theory
on $AdS_5 \times S^5$ and conformally invariant Yang-Mills theories
~\cite{Gubser:1998bc,Witten:1998qj} is the potential to describe
processes for physical QCD which are valid at strong coupling and do
not rely on perturbation theory. As shown by  Polchinski and
Strassler~\cite{Polchinski:2001tt}, dimensional counting
rules~\cite{Brodsky:1973kr} for the leading power-law fall-off of
hard exclusive  scattering can be derived from a gauge theory with a
mass gap dual to supergravity in warped spacetimes. The modified
theory generates the hard behavior expected from QCD, instead of the
soft behavior characteristic of strings. Other examples are the
description of form factors at large transverse
momentum~\cite{Polchinski:2001ju} and deep inelastic
scattering~\cite{Polchinski:2002jw}. The discussion of scaling laws
in warped backgrounds has also been addressed
in~\cite{Boschi-Filho:2002zs,Brower:2002er,Andreev:2002aw}.

The AdS/CFT correspondence has now provided important new
information on the short-distance structure of hadronic LFWFs; one
obtains conformal constraints which are not dependent on
perturbation theory. The large $k_\perp$ fall-off of the valence
LFWFs is also rigorously determined by consistency with the
evolution equations for the hadron distribution
amplitudes~\cite{Lepage:1980fj}. Similarly, one can also use the
structure of the evolution equations to constrain the $x \to 1$
endpoint behavior of the LFWFs. One can use these strong constraints
on the large $k_\perp$ and $x \to 1$ behavior to model the LFWFs.
Such forms can also be used as the initial approximations to the
wavefunctions needed for variational methods which minimize the
expectation value of the light-front Hamiltonian. The derivation is
carried out in terms of the lowest dimensions of interpolating
fields near the boundary of AdS, treating the boundary values of the
string states $\Psi(x,r)$ as a product of quantized operators which
create $n$-partonic states out of the vacuum~\cite{Brodsky:2003px}.
The AdS/CFT derivation validate QCD perturbative results and confirm
the dominance of the quark interchange
mechanism~\cite{Gunion:1972gy} for exclusive QCD processes at large
$N_C$. The predicted orbital dependence coincides with the fall-off
of light-front Fock wavefunctions derived in perturbative
QCD~\cite{Ji:2003fw}. Since all of the Fock states of the LFWF
beyond the valence state are a manifestation of quantum
fluctuations, it is natural to match quanta to quanta the additional
dimensions with the metric fluctuations of the bulk geometry about
the fixed AdS background. For example, the quantum numbers of each
baryon, including intrinsic spin and orbital angular momentum, are
determined by matching the dimensions of the string modes
$\Psi(x,r)$, with the lowest dimension of the baryonic interpolating
operators in the conformal limit.

\begin{figure}
\centering
\includegraphics[angle=0,width=9cm]{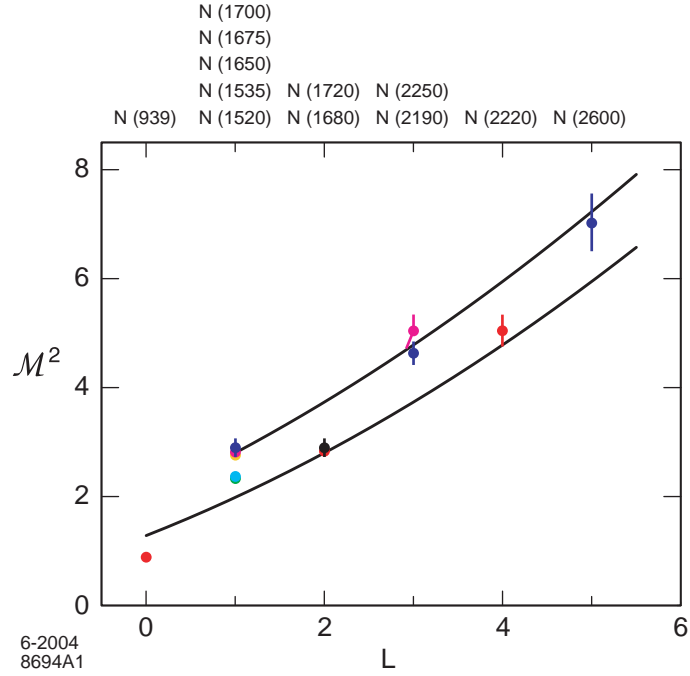}
\caption{Nucleon orbital spectrum for a value of $\Lambda_{QCD}$ = 0.22 GeV.
The lower curve corresponds to nucleon
  states dual to spin-$\half$ string modes in the bulk. The upper curve corresponds to nucleon
states dual to string-$\threehalf$ modes.}
\label{fig:Nucleonspec}
\end{figure}

\begin{figure}
\centering
\includegraphics[angle=0,width=9cm]{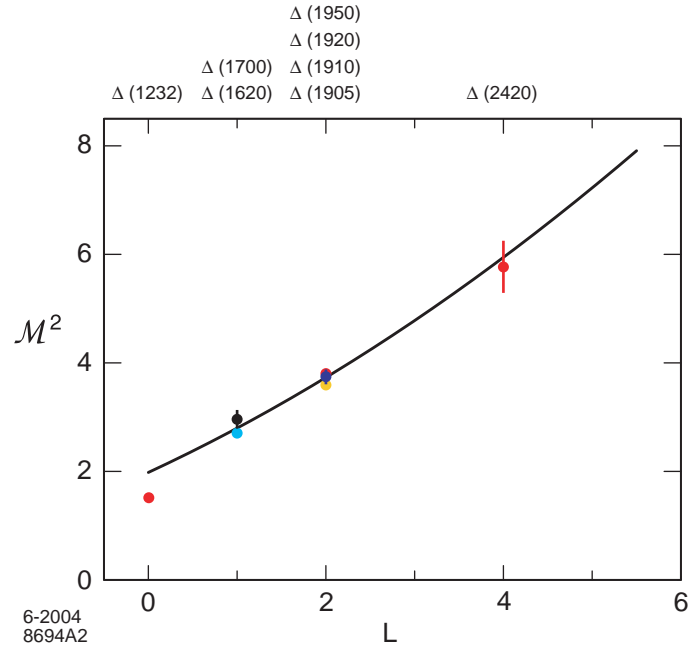}
\caption{Delta orbital spectrum for $\Lambda_{QCD}$ = 0.22 GeV.
 The Delta states dual to spin-$\half$ and
 spin-$\threehalf$ string modes in the bulk lie on the same
 trajectory.}
\label{fig:Deltaspec}
\end{figure}

The AdS/CFT correspondence also provides a novel way to compute the hadronic spectrum.
The essential assumption is to require the hadron wavefunctions to vanish at the
fifth-dimensional coordinate $r_0= \Lambda_{QCD} .$ As an example,  Fig.
\ref{fig:Nucleonspec} shows the orbital spectrum of the nucleon states and in Fig.
\ref{fig:Deltaspec}\ the $\Delta$ orbital resonances recently computed by Guy de Teramond
and myself~\cite{deTeramond:2004qd}. The values of $\mathcal{M}^2$ are computed as a
function of orbital angular momentum $L$. The nucleon states with intrinsic spin $S =
\half$ lie on a curve below the nucleons with $S = \threehalf$. We have chosen our
boundary conditions by imposing the condition $\Psi^+(x, z_o) = 0$ on the positive
chirality modes for $S = \half$ nucleons, and $\Psi^-_\mu(x, z_o) = 0$ on the chirality
minus strings for $S =\threehalf$. In contrast to the nucleons, all of the know $\Delta$
orbital states with $S = \half$ and $S = \threehalf$ lie on the same trajectory. The
boundary conditions in this case are imposed on the chirality minus string modes. The
numerical solution corresponding to the roots of Bessel functions give the nonlinear
trajectories indicated in the figures. All the curves correspond to the value
$\Lambda_{QCD} = 0.22$ GeV, which is the only actual parameter aside from the choice of
the boundary conditions. The results for each trajectory show a clustering of states with
the same orbital $L$, consistent with strongly suppressed spin-orbit forces; this is a
severe problem for QCD models using one-gluon exchange. The results also indicate a
parity degeneracy between states in the parallel trajectories shown in Fig.
\ref{fig:Nucleonspec}, as seen by displacing the upper curve by one unit of $L$ to the
right. Nucleon states with $S = \threehalf$ and $\Delta$ resonances fall on the same
trajectory~\cite{Klempt:2004yz}.

Since only one parameter, the QCD scale $\Lambda_{QCD}$, is used,
the agreement of the model with the pattern of the physical light
baryon spectrum is remarkable. This agreement possibly reflects the
fact that our analysis is based on a conformal template, which is a
good initial approximation to QCD~\cite{Brodsky:2004qb}. We have
chosen a special color representation to construct a three-quark
baryon, and the results are effectively independent of $N_C$. The
gauge/string correspondence  appears to be a powerful organizing
principle to classify and compute the mass eigenvalues of baryon
resonances.

\bigskip

\section*{Acknowledgments}

I wish to thank Guenther Rosner,  Stanislav Belostotski, David
Ireland, Douglas MacGregor and their colleagues at  Glasgow
University and Edinburgh University for organizing the 2004 Scottish
Universities Summer School in Physics at St. Andrews.  These
lectures are based on collaborations with Carl Carlson, Guy de
Teramond, Markus Diehl, Rikard Enberg, John Hiller, Kent Hornbostel,
Paul Hoyer, Dae Sung Hwang, Gunnar Ingelman, Chueng Ji,  Volodya
Karmanov,  Peter Lepage, Gary McCartor, Chris Pauli, Joerg
Raufeisen,  Johan Rathsman, and Dave Robertson.

\end{document}